\documentclass[]{elsarticle}

\usepackage[ruled]{algorithm2e}

\SetAlFnt{\small}
\SetAlCapFnt{\small}
\SetAlCapNameFnt{\small}
\SetAlCapHSkip{0pt}
\IncMargin{-\parindent}

\usepackage{tikz}
\def\checkmark{\tikz\fill[scale=0.4](0,.35) -- (.25,0) -- (1,.7) -- (.25,.15) -- cycle;}

\usepackage[normalem]{ulem}
\usepackage{url}
\usepackage{graphicx}
\usepackage[english]{babel}
\usepackage{amsmath}
\usepackage{subcaption}
\usepackage{wrapfig}
\usepackage{multicol}
\usepackage{caption}
\usepackage{amssymb}
\usepackage{lipsum}

\usepackage[colorinlistoftodos]{todonotes}
\usepackage[colorlinks = true,
            linkcolor = Maroon,
            urlcolor  = NavyBlue,
            citecolor = Maroon,
            anchorcolor = NavyBlue,
            bookmarks=false]{hyperref}

\usepackage{arydshln}

\usepackage{soul}

\usepackage{multirow}
\usepackage{booktabs}
\usepackage{pifont}
\newcommand*\rot{\rotatebox{90}}
\usepackage{pbox}

\usepackage{forest}
\usetikzlibrary{trees}

\usepackage{rotating}

\journal{}

\begin{document}
\begin{frontmatter}
\title{Mobile Cloud Business Process Management System for the Internet of Things: A Survey}

\author[ut1]{Chii Chang\corref{mycorrespondingauthor}}
\cortext[mycorrespondingauthor]{Corresponding author}
\ead{chii.chang@acm.org}

\author[ut1]{Satish Narayana Srirama}

\author[um1]{Rajkumar Buyya}

\address[ut1]{University of Tartu, Estonia}
\address[um1]{University of Melbourne, Australia}

\begin{abstract}
The Internet of Things (IoT) represents a comprehensive environment that consists of a large number of smart devices interconnecting heterogeneous physical objects to the Internet. Many domains such as logistics, manufacturing, agriculture, urban computing, home automation, ambient assisted living  and various ubiquitous computing applications have utilised IoT technologies. Meanwhile, Business Process Management Systems (BPMS) have become a successful and efficient solution for coordinated management and optimised utilisation of resources/entities. However, past BPMS have not considered many issues they will face in managing large scale connected heterogeneous IoT entities. Without fully understanding the behaviour, capability and state of the IoT entities, the BPMS can fail to manage the IoT integrated information systems. In this paper, we analyse existing BPMS for IoT and identify the limitations and their drawbacks based on Mobile Cloud Computing perspective. Later, we discuss a number of open challenges in BPMS for IoT.

\end{abstract}

\begin{keyword}
Internet of Things, Mobile Cloud Computing, Business Process Management System, Service-Oriented, review, state of the art, challenge.
\end{keyword}

\end{frontmatter}

\section{Introduction}
Emerging mature mobile and ubiquitous computing technology is hastening the realisation of smart environments, in which the physical objects involved in our everyday life (food, parcels, appliances, vehicles, buildings etc.) are connected. Many of the electronic devices are now granted with a certain intelligence to work together for us and enhance our life. The core enabler is the Internet Protocol (IP) that is capable of providing the addressing mechanism for physical objects towards interconnecting everything with the Internet, which is known as the Internet of Things (IoT) \cite{Gubbi2013}.~
The goal of IoT is to enhance the broad range of people's lives, including but not limited to agriculture, transportation, logistics, education, healthcare and more \cite{Atzori2010}. The industry predicts in the year 2020, around 50 billion physical devices will be connected to the Internet \cite{Evans2011}, and the economy revenue value will raise up to \$1.9---\$7.1 trillion \cite{Middleton2013,Manyika2013,Lund2014}.

Today, IoT has become one of the most popular topics for both industry and academia.~
Prior to the vision of IoT has arisen, the Cyber-Physical Systems (CPS), which interconnected the physical entities with software systems, are usually isolated. Each sub-system has its own topology and communication protocols. In order to integrate the isolated CPS into the IoT vision, one promising approach is to apply Service-Oriented Architecture (SOA)-based middleware solution \cite{Loke2003}.~
Fundamentally, SOA introduces the interoperability of heterogeneous isolated systems. By Applying SOA, CPS can manage individual devices as atomic services or they can configure a group of devices to provide composite services. For example, a mobile Internet-connected device can embed a Web service to provide the atomic sensory information service to remote Web service clients \cite{Srirama2006,Chang2015maasym}.~
Further, a composite service in the cloud can exploit the data derived from multiple devices in a specific location to compute and generate the meaningful information to specific users \cite{Conti2012}.

In the last decade, Workflow Management Systems (WfMS) have become one of the major components of service composition. Among the different WfMS, Business Process Management Systems (BPMS) have been broadly accepted the de facto approaches in WfMS \cite{Dumas2013}, mainly because of the availability of tools and international standards such as Business Process Execution Language (BPEL) \cite{Jordan2007BPEL}, Business Process Model and Notation (BPMN) \cite{Model2011BPMN} and XML Process Definition Language (XPDL) \cite{WfMC2012}. BPMS are~
{\it ``generic software system(s) that is driven by explicit process designs to enact and manage operational business processes"} \cite{Aalst2003}.~
BPMS can provide the highly integrated platforms to manage comprehensive entities and activities involved in IoT systems. BPMS also provide self-managed behaviour in various IoT applications such as smart home system \cite{Loke2003,Chang2008}, crowd computing \cite{Kucherbaev2013}, Wireless Sensor Network \cite{Sungur2013}, heating, ventilating and air conditioning (HVAC) system \cite{Tranquillini2012}, mobile healthcare \cite{Peng2014} and so on.~
BPMS let users (e.g. system administrators, domain scientists, regular end users) to easily manage the overall IoT system without getting involved in the low-level complex programming languages. Hence, they can focus on modelling the behaviour and business processes of the things. Furthermore, with the optimised process model, the IoT system can provide self-adaptation in which it can autonomously react to or prevent the events based on runtime context information \cite{Chandler2015}.~
Below, we summarise a few use cases of IoT-driven BPMS.
\vspace*{-5pt}
\subsection{Use Cases of IoT-driven Business Process Management Systems}\label{sec:usecases}
\noindent \textbf{Logistic.}~
By integrating mobile cloud, IoT and BPMS, the logistics system can provide real-time tracking and controlling. For example,~ imagine a cargo parking area needs to avoid the cargos with dangerous goods to be parked in close proximity. By deploying Radio-Frequency IDentification (RFID), wireless sensors and mobile ad hoc network at the front-end vehicles, the front-end vehicles can maintain a temporary edge network to identify the goods in their cargos and to inform each other if their driver intends to park the vehicles in close proximity.~
The distant cloud-side management system can continuously track the environment and provide the instructions to the vehicle drivers regarding where the proper free space for parking the vehicles and cargos is \cite{Glombitza2011}.

\smallskip
\noindent 
\textbf{Enterprise Resource Planning (ERP).}~
The supply chain process in  ERP systems is another example that can be improved by BPMS for IoT (BPMS4IoT). As described in \cite{Schulte2014}, by implementing wireless sensor networks, RFID, or other IoT technologies, the supply chain can be monitored in real time. Suppose Factory-A has a production lane which requires the supply from Factory-B urgently.~
Since IoT has been deployed in all the Factory-A's partner businesses, Factory-A is capable of identifying whether Factory-B can produce and ship the supply to them in time or not based on analysing Factory-B's in-stock resources and shipping conditions. In the case of Factory-A found that Factory-B is unable to fulfil the task in time due to the shipping issue, Factory-A can try to find a substitution by either distributing the order to multiple suppliers or by finding an alternative supplier which can handle the entire order.~
If Factory-A is unable to find an alternative solution and it has to postpone the production lane, the originally assigned workers for production lane and the vehicles for shipment can be re-allocated to the other tasks in order to reduce the waste of human resources.

\medskip
\noindent 
\textbf{Smart Building.}~
BPMS4IoT can improve the efficiency of people's everyday living areas. BPMS enables autonomous actions to be triggered based on certain events that are measured based on the deployed front-end IoT devices. For example, the heating, ventilating, and air conditioning (HVAC) systems used in the modern buildings can be monitored and controlled from the remote BPM-based information system precisely via Internet-connected wireless sensor and actuator network (WSAN) devices.~
A simple smart home system that integrates with the mobile service can identify the house resident's movement (via their mobile/wearable devices) in order to maintain the oxygen and temperature level when the resident is coming back home from the work. The indoor monitoring devices also can track the home environment and inform any event occurred to the resident remotely when he or she is away from home.~
For the large residential building such as the hotel HVAC control system, the BPMS4IoT enhances the efficiency of the management in which the system can automatically measure the usage of  electricity and heating system and bill the customer based on the usage instead of charging the fixed rate for each room \cite{Tranquillini2012}.

\medskip
\noindent 
\textbf{Health Care and Ambient Assisted Living (AAL).}~
The BPMS4IoT-based health care system integrates cloud services with Wireless Body Area Network (WBAN) \cite{Quwaider2015}, which is formed by numerous wearable sensor devices that can measure blood sugar, body temperature, heartbeat etc. We take the use case described in \cite{Dar2011} as an example.

Imagine a 70 years old woman---Emily, who is using the remote Eldercare system with WBAN attached to her body to measure her blood sugar level. One day, she feels a small headache and dizziness. The sensor has detected that Emily's blood sugar has violated the predefined threshold. Hence, the report is immediately sent to the hospital's system to inform the physician. The remote physician then performs the remote health monitoring process via the Internet-connected body sensor network attached with Emily.~
Meanwhile, the system also informs Emily's son via SMS about Emily's health condition. Afterward, Emily receives the prescription and dietary recommendations from the physician. A while later, Emily's son visits her and assist Emily to recover her health condition back to normal \cite{Dar2011}.

\subsection{Problem Statement}
Although many IoT application domains have applied BPMS and they have shown the promising solutions, many of them have not considered the challenges that the BPMS will face in the near future IoT environments. We summarise the challenges below:

\begin{figure*}[h]
  \centering
    \includegraphics[width=0.95\textwidth]{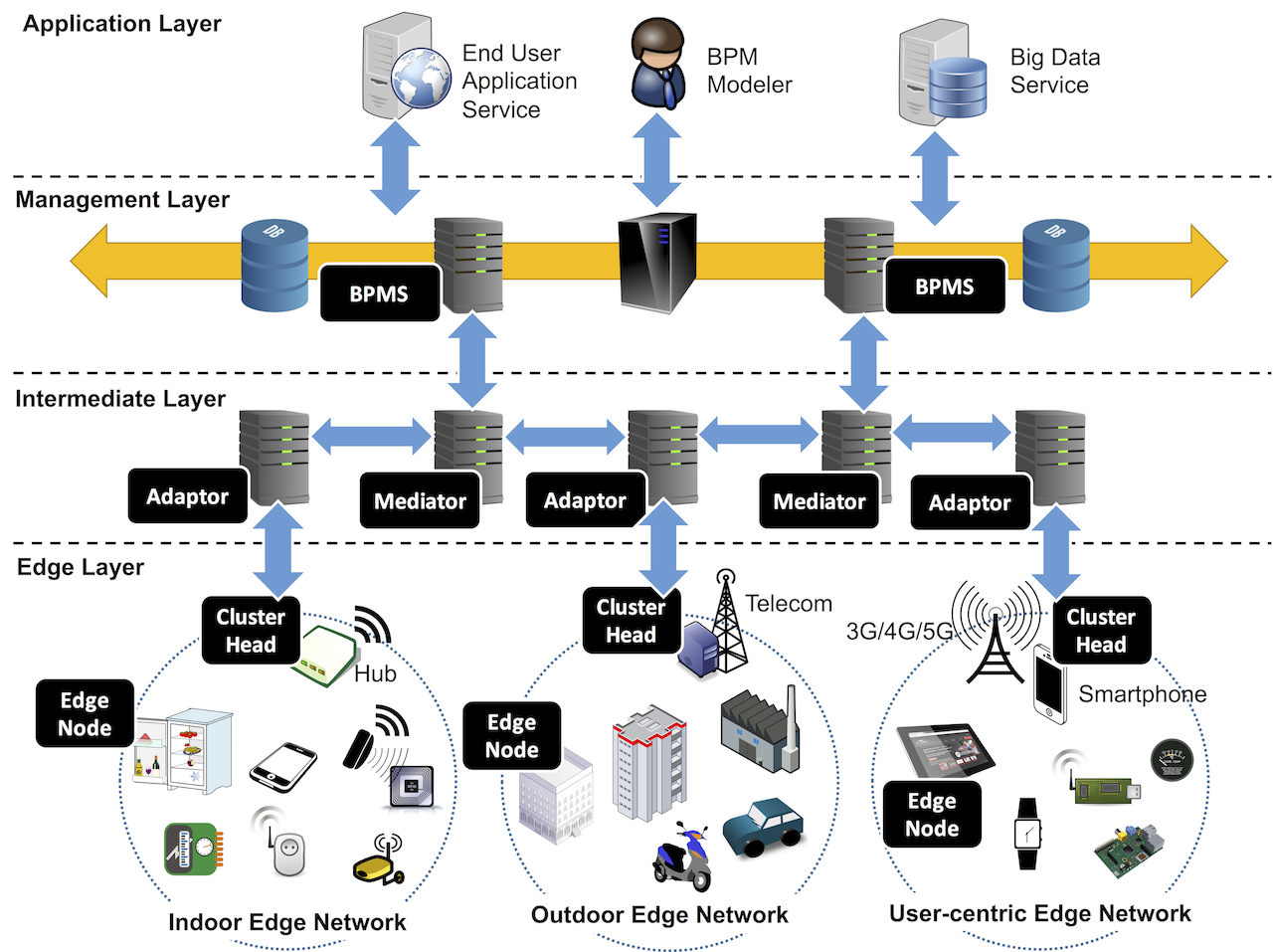}
    \caption{Simplified common SOA-based BPMS4IoT.}
    \label{fig_current_architecture}
\end{figure*}
Fig.~\ref{fig_current_architecture} illustrates a common SOA-based BPMS architecture that involves IoT entities and the middleware-based information system. Most BPMS4IoT frameworks are following the design in the figure. Such a design faces the following limitations.

%

\begin{itemize}
\item \textit{Transparency}. As the Figure shows, the system has no direct interaction with the edge network (i.e. front-end wireless Internet-connected nodes such as mobile phones, wireless sensors, data collectors etc.).~
The design that hides the detail topology and communication at the edge network may cause the process designer unable to properly define the behaviours and processes of the devices in terms of adding participants, modifying or customising the processes, react to or prevent failures and so on.~
Further, recent IoT management systems often involve edge nodes as a part of the BPMS in the scenarios such as logistics \cite{Glombitza2011}, crowdsourcing processes \cite{Tranquillini2015}, HVAC system \cite{Tranquillini2012}, health care services \cite{Peng2014}, Ambient Assisted Living \cite{Dar2015}, real-time sensing \cite{Chang2015scc} where mobile and wireless network objects are involved.
\item \textit{Agility}. SOA-based IT systems often use interface and middleware technologies to leverage different entities. However, due to the lack of standardisation in process modelling and execution levels, systems usually resulted in a complex and inflexible design.~
For example, based on the architecture described in Figure~\ref{fig_current_architecture}, if two devices, which are connected to different backend servers, but located in physical proximity, intend to interact with each other, the communication needs to pass through multiple layers, in which the direct interaction does not exist and it also affects the overall performance.~
As the vision \cite{Conti2012} indicates, in the future IoT environments, devices will cooperate automatically for certain tasks in order to improve the efficiency and agility.
\end{itemize}

\subsection{Aim of the Survey}
In this paper, we review the existing BPMS for IoT (BPMS4IoT)  frameworks to identify whether or not they have addressed the challenges in BPMS4IoT and how they overcome the challenges from the perspective of Mobile Cloud Computing (MCC).

MCC represents the integration of mobile computing and cloud computing in which it addresses the elastic cloud resource provisioning and the optimised interaction between cloud services and mobile network nodes. Commonly, disciplines in MCC addressed many studies in mobile connectivity, mobility, discovery, resource-awareness, decentralised service interaction and how the system efficiently integrates the mobile entities as process participants with cloud services.

In the past several years, researchers have proposed numerous MCC-related WfMS, including the mobile device-hosted WfMS engines \cite{Hackmann2006,Pryss2011,Sen2008,Chou2009}, WfMS for enabling mobile ad-hoc cloud computing \cite{Chang2014mum}, WfMS for IoT that are exploiting both mobile and utility cloud resources \cite{Chang2012icsoc,Chang2014pmc,Chang2015scc}. Therefore, MCC has shown the promising solution for the efficient integration of information systems with various wireless Internet-connected entities for distributed process management. Furthermore, the deployments of existing IoT management systems are widely utilising wireless sensor network, mobile and cloud services, which indicates that MCC is the key enabler of BPMS4IoT. Therefore, in this paper, we intend to address the issues of BPMS4IoT by applying MCC concepts.

The rest of the paper is organised as follows: Section 2 provides a brief review of related literature surveys.~
Section 3 provides the state-of-the-art literature review on BPMS4IoT frameworks based on the life cycle phases of BPMS4IoT and Section 4 compares the existing frameworks. In Section 5 we identify the open challenges and issues that have not yet been fully addressed in existing works. The paper concludes in Section 6.

\section{Related Work}

There exist a large number of related literature surveys in IoT and BPMS. In this section, we summarise the featured works with categorisation below.
\begin{itemize}
\item {\it Comprehensive}. 
Numerous surveys \cite{Gubbi2013,Atzori2010,Vermesan2011,Miorandi2012,Borgia2014,Guo2013,DaXu2014} provide the comprehensive review of existing IoT and related works. They are focused on discussing the background of IoT, emerging technologies, promising applications in various domains and open challenges.
%
\item {\it Service discovery}. 
The term---{\it service} in IoT domain represents the function provided by the CPS that is interconnected with the physical entities. Considering a large number of heterogeneous physical entities will be connected to the Internet in the near future, service discovery becomes extremely challenging.~
Especially in the big data service environment in which the data is retrieved from various spatiotemporal service providers. Therefore, a number of literature survey papers \cite{Evdokimov2010,Villaverde2014} are focused on the service discovery in IoT.
%
\item {\it Network technology}. 
Since the vision of IoT been introduced, researchers have acknowledged the importance of feasible network communication protocols for resource constrained devices, which are the core front-end elements of IoT. Numerous literature surveys \cite{Mainetti2011,Palattella2013,Sheng2013,Tonneau2015} are focused on reviewing the feasibility of the existing and emerging network protocols for IoT.
%
\item {\it Middleware}. 
Middleware technology is the enabler to integrate the front-end physical entities with the back-end software systems. Numerous literature surveys are focused on reviewing and comparing existing middleware technologies for IoT. These include an overview of the existing CPS middleware framework \cite{Bandyopadhyay2011,Bandyopadhyay2011a,Chaqfeh2012}, a review on Web service-oriented IoT middleware \cite{Zeng2011,Issarny2011}, cloud computing and IoT integration \cite{Botta2016} and a study on how to select proper protocols for the connected devices~\cite{Mashal2015}.
%
\item {\it Domain specific}. 
There also exists a number of IoT literature surveys that focus on specific research area such as health care and disabilities \cite{Domingo2012,Islam2015}, urban computing \cite{Zanella2014,Conti2012,Salim2015}, multimedia \cite{Alvi2015},~
data mining  and big data \cite{Aggarwal2013,Tsai2014}, social network aspect \cite{Atzori2012}, energy efficiency \cite{Villaverde2012,Aziz2013}, mobility \cite{Zorzi2010,Silva2014,Bouaziz2014}, trust management \cite{Yan2014} and security \cite{Roman2013,Sicari2015,Nguyen2015,Granjal2015}. All of them have provided a detailed study on the specific domain when IoT is applied. 

\item \textit{Business Process Management}. Numerous of BPM-based survey papers have been published in recent years. These papers include but not limit to the comprehensive survey \cite{Aalst2013} that overviews the process modelling methods, techniques and tools, business process modelling standards \cite{Ko2009}, elastic cloud system-driven BPM \cite{Schulte2015} and business process model frameworks \cite{Yan2012}.

\end{itemize}

Although there exist a large number of literature survey papers in the domain of IoT and BPMS today, to the best of our knowledge, this is the first literature survey that focuses on BPMS4IoT in MCC perspective. BPMS4IoT faces its specific challenges that have not yet been addressed in the past IoT survey papers.


\section{Integrating BPMS with IoT}

In general, we can consider three phases in the life cycle of BPMS:   {\it (re)design phase}, {\it implement/configure phase} and {\it run and adjust phase} \cite{Aalst2013}. Depending on the focus, some disciplines can further classify each phase into the more detailed phases. For example, the authors in their earlier works \cite{Aalst2003,Weske2004} have separated the {\it (re)design phase} to {\it diagnosis phase} and {\it process design phase}. The {\it (re)design phase} can also involve three different phases, including {\it process discovery}, {\it process analysis} and {\it process redesign} \cite{Dumas2013}. Although there can be other notions of defining the life cycle, in this paper, we apply the recent definition of BPM life cycle from van der Aalst et al. \citeyear{Aalst2013} to BPMS4IoT.

\begin{figure}[h]
  \centering
    \includegraphics[width=0.35\textwidth]{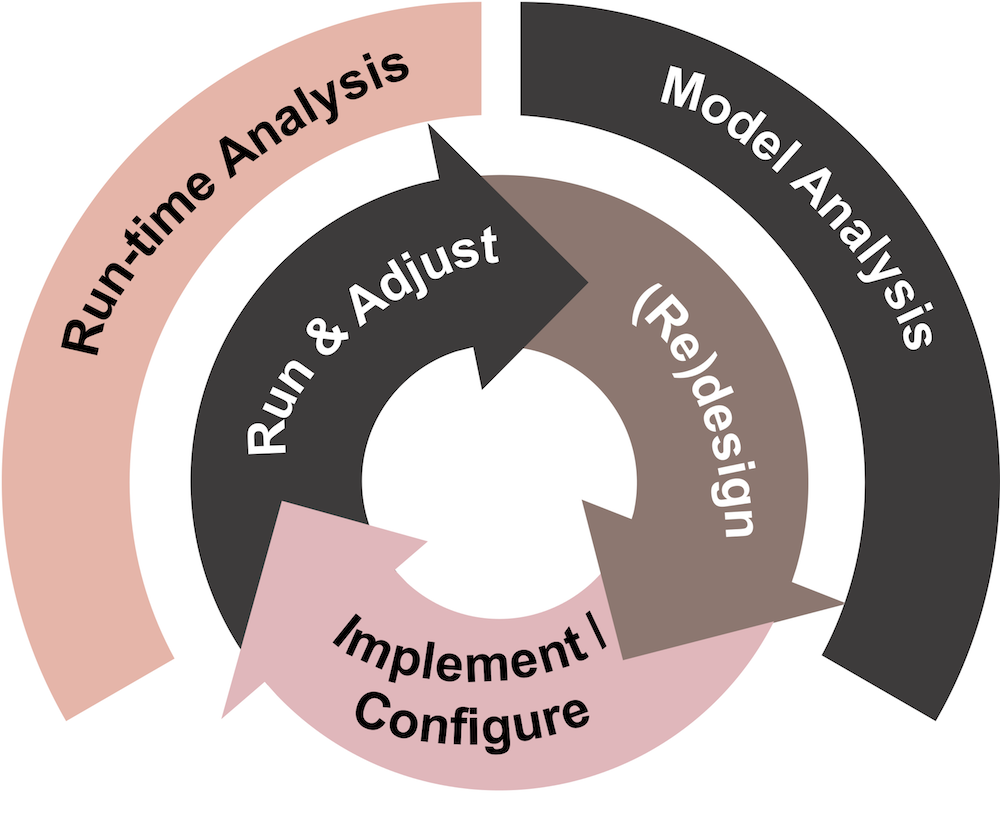}
    \caption{BPMS4IoT life cycle.}
    \label{fig_bpm_lifecycle}
\end{figure}

Figure~\ref{fig_bpm_lifecycle} illustrates the life cycle of BPMS4IoT. The {\it (re)design phase} involves how to model the connected IoT elements, their related elements and how their behaviour is in the business process. The {\it implement/configure phase} involves how to practically implement the process model as executable methods. Further, it involves how to deploy the executable methods to the corresponding workflow engines for execution. The {\it run and adjust phase} corresponds to how the BPMS autonomously monitors and manages the system at runtime, and how it continuously improves and optimises the process. If the process designer recognises the need of improving the processes, they will re-designed the system and continue with the cycle. 

Aside from the three main life cycle phases, during the {\it (re)design phase}, the system developers will perform technologies to analyse the model design (e.g. using simulation). The management team also collects the data (e.g. event logs) in the {\it run and adjust phase} for process diagnosis.

\medskip

The following subsections provide a state-of-the-art review and analysis based on each phase of the life cycle.

\subsection{(Re)design Phase}\label{section_design}
The {\it (re)design} phase of BPMS4IoT involves: 

\begin{enumerate}
\item {\it Architecture}---represents the fundamental design of the system,  which can base on the centralised orchestration model, the decentralised choreography model or the hybrid model that inherits the features from both.
\item {\it Modelling}---involves how to model the business processes. The process model will use only existing methods (e.g. notations of standard BPMN) to design the IoT entities and their activities or will it introduce new elements for best describing the model? further, it involves what entities need to be considered?
\item {\it Transparency}---indicates that the process modelling should provide a comprehensive view of the overall execution environment. 
\end{enumerate}

Below, we discuss the subjects that have been addressed in the existing BPMS4IoT frameworks for the {\it (re)design phase}.

%
%

%

\subsubsection{Orchestration vs. Choreography}
%
%

BPMS can be broadly classified into two types: orchestration and choreography. The orchestration is mainly based on a centralised architecture in which a single management system is managing the entire process execution. On the other hand, choreography represents a system that 
in certain stages, a number of external information systems are handling the processes (or the portions of the process). While BPMS4IoT is commonly designed based on orchestration model,~
\begin{figure*}[h]
    \centering
    \begin{subfigure}[t]{0.45\textwidth}
        \includegraphics[width=\textwidth]{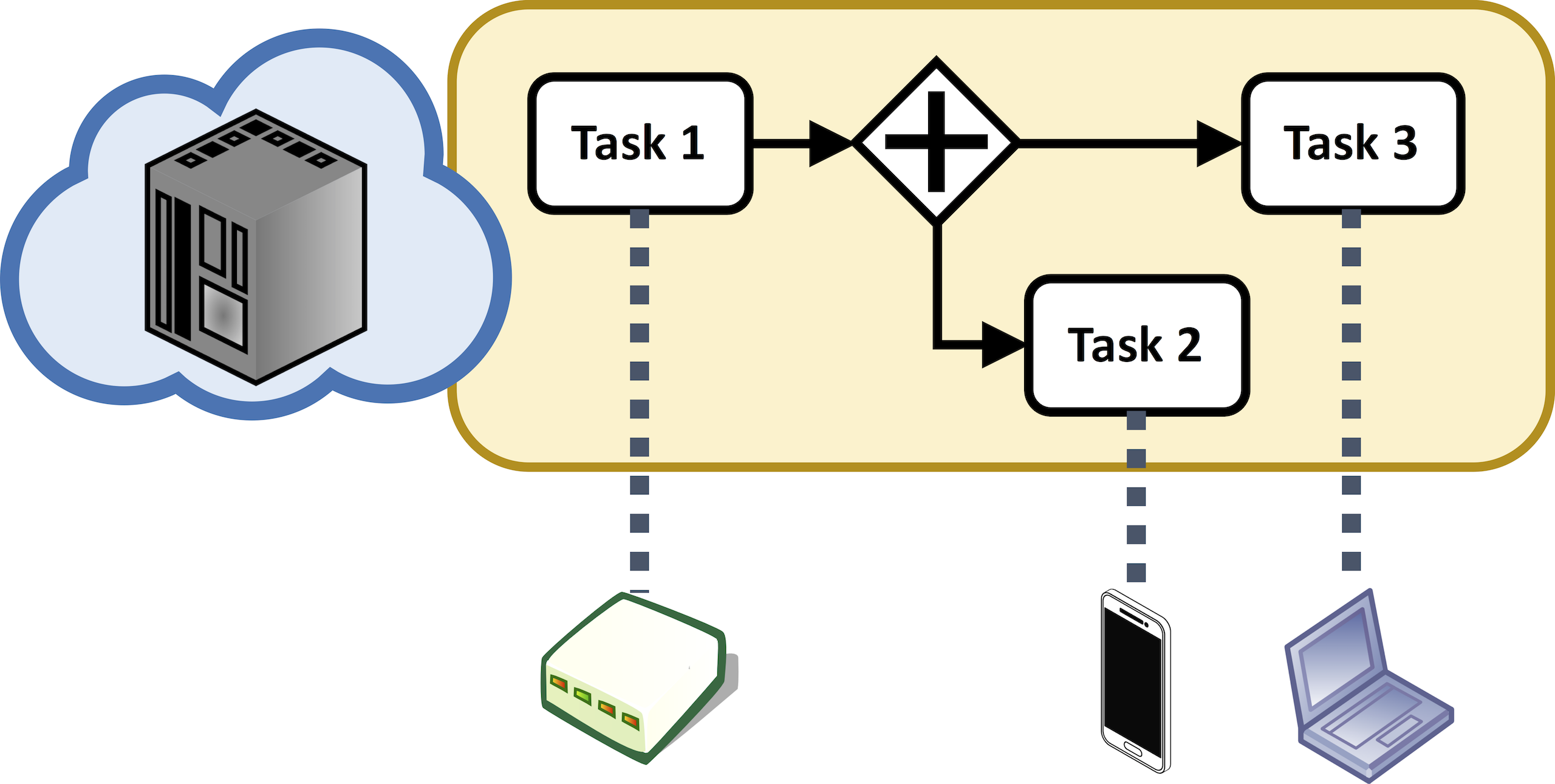}
        \caption{Centralised orchestration}
        \label{fig:orchestration}
    \end{subfigure}~~~~~~
    \begin{subfigure}[t]{0.45\textwidth}
        \includegraphics[width=\textwidth]{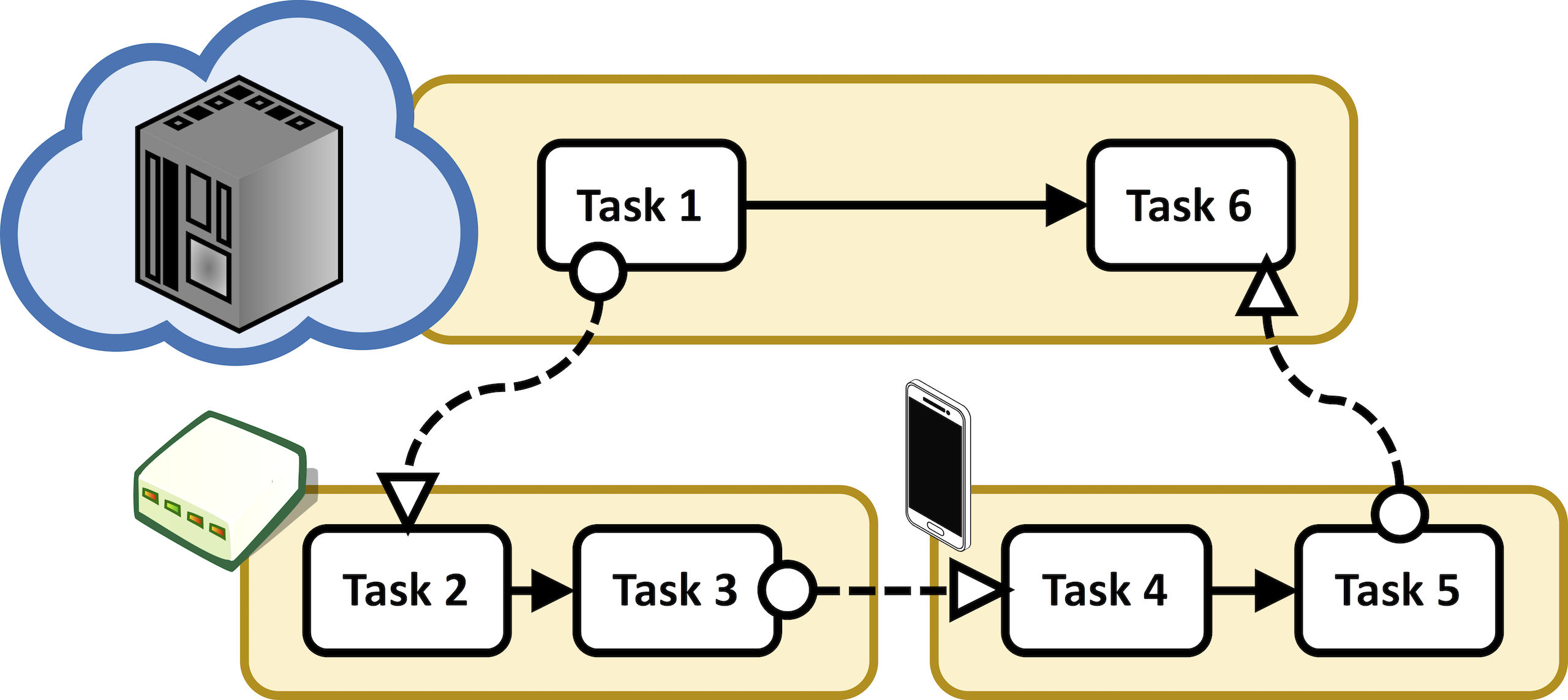}
        \caption{Decentralised choreography}
        \label{fig:choreography}
    \end{subfigure}
    \caption{Business process execution: orchestration vs. choreography.}\label{fig:orch_vs_chore}
\end{figure*}
recent works have emphasised the importance of choreography in IoT. For example, Dar et al. \cite{Dar2011,Dar2015} considered that the centralised orchestration is insufficient to agilely react to events occurring on the edge network. Especially in an inter-organisational network or in the mobile IoT scenarios, which involve numerous mobile Internet-connected participants.~

In such scenarios, there is a need in applying choreography-based architecture to enable a certain degree of business process distribution, in which the edge nodes will need the Business Process (BP) model execution mechanisms and self-management abilities. Moreover, distributing process execution at edge nodes can further enhance the flexibility, agility and adaptability of the BPMS4IoT \cite{Dar2015,Peng2014,Tranquillini2012}.

\subsubsection{Existing vs. Extension}
%
Introducing the IoT elements in BPMS is not a straightforward task because IoT devices can be heterogeneous in terms of communication protocols, network topologies (e.g. connectivity), hardware specifications (e.g. computing power and battery life). In general, there are two approaches to model the entities of IoT:

\begin{enumerate}
\item \textit{Expressing IoT devices as services}.~ 
%
Modelling the IoT devices as URI-based services can simplify the management systems and also be fully compatible with existing tools such as BPMN or BPEL. In this approach, the system expresses IoT devices as regular network services such as XML-based W3C---SOAP Web services \cite{Peng2014}, OASIS---Devices Profile for Web Services (DPWS) or OGC Web Service Common, which is communicable via the regular request-response methods. Generally, in this approach, BP model designers assumed that the system can connect to IoT devices directly based on the Web service communication. However, in real world systems, many IoT devices do not work as regular Web service entities.  Therefore, numerous researchers are focusing on defining new elements for BPMS4IoT.

\item \textit{Defining new IoT elements}.~
%
In general, BPMS such as Enterprise Resource Planning (ERP) systems often assumed that the system will provide automation for all the involved devices, and the system will have the capability to directly invoke all the devices. However, in IoT systems, such assumption, in many cases, is not applicable \cite{Meyer2015}. IoT entities that have different capabilities may connect with the management system differently. 

Beside the regular direct IP network connected devices, in many scenarios, the IoT devices connect to the management system via multiple network layers or routings. For example, an actuator device may connect with an intermediary service provided by different devices in order to let the management system access it. 

Another example is when the BPMS involves continuous tasks such as sensor data streaming and Eventing, existing standard-based BPM modelling tools cannot explicitly define the process \cite{Appel2014}. In such cases, the system needs a more proper way to let designers to model the process accordingly in order to fully manage and optimise the systems. Thus, a common approach is to introduce specific IoT elements in BPM to differentiate them from the traditional BPM elements such as service tasks in BPMN.

\end{enumerate}


\subsubsection{Modelling IoT Elements}\label{sec:modellingThings}
%
As mentioned previously, modelling IoT elements in BPM is not a straightforward task. Although designers can extend the business process modelling languages such as standard BPMN 2.0 and the corresponding tools to introduce the IoT elements by either expressing them as components or simply describing them as Uniform Resource Identifier (URI)-based services, overall, existing BPM solutions have not included IoT specifically \cite{Meyer2015}. Consequently, numerous related works tended to introduce new notations that represent IoT elements in the business process models.

Table~\ref{table_IoTElements} summarises the IoT elements  modelled in the existing BPMS4IoT frameworks.

\begin{table*}[h]
\scriptsize
\centering
\caption{
BPMN extension for introducing IoT
}
\label{table_IoTElements}
\begin{tabular}{l|llllllll}
\hline
Project 
& \rot{\shortstack[l]{IoT System\\Processes}}
& \rot{\shortstack[l]{IoT Device\\Activity}}
& \rot{\shortstack[l]{Physical\\Entity}} 
& \rot{Sensor} 
& \rot{Actuator} 
& \rot{\shortstack[l]{Intermediary\\Operation}}
& \rot{Event Stream} 
& \rot{\shortstack[l]{Special\\Element}}
\\
\hline
IoT-A       
& PO($\downarrow$) 
& LN($\downarrow$) 
& PA($\uparrow$) 
& TA($\sim$) 
& TA($\sim$) 
&   
&  
&
\\ \hdashline[1pt/2pt]
Sungur et al.
& PO($\sim$) 
&
&  
& ST($\sim$) 
& ST($\sim$) 
& ST($\uparrow$) 
&  
& 
\\ \hdashline[1pt/2pt]
uBPMN
&  
&   
&   
& \pbox{20cm}{TA/EV($\uparrow$)} 
& TA/EV($\uparrow$)  
&
&  
& DO($\uparrow$)
\\ \hdashline[1pt/2pt]
SPUs
& 
&
& 
&  
&   
&    
& \pbox{20cm}{DO/\\ST($\uparrow$)} 
& 
\\ \hdashline[1pt/2pt]
makeSense
& PO($\uparrow$) 
&  
&  
& TA($\sim$)
& 
& 
& 
& 
\\
\hline
\multicolumn{9}{l}{Subject addressed level: ($\uparrow$) = high; ($\sim$) = medium; ($\downarrow$) = low;}
\\
\end{tabular}
\begin{tabular}{llllllll}
\\
\hline
PO & BPMN Pool   & PA & Participant  & TA & BPMN Task 
& DO & Data Object        
\\
LN & BPMN Lane   & EV & BPMN Event   & ST & BPMN Service Task
\\ \hline
\end{tabular}
\end{table*}

As the table shows, almost all the modelling frameworks have introduced sensor element. Followed by the actuator and the IoT System Processes.  We summarise each element below.


\medskip
\begin{figure}[h]
  \centering
    \includegraphics[width=1\textwidth]{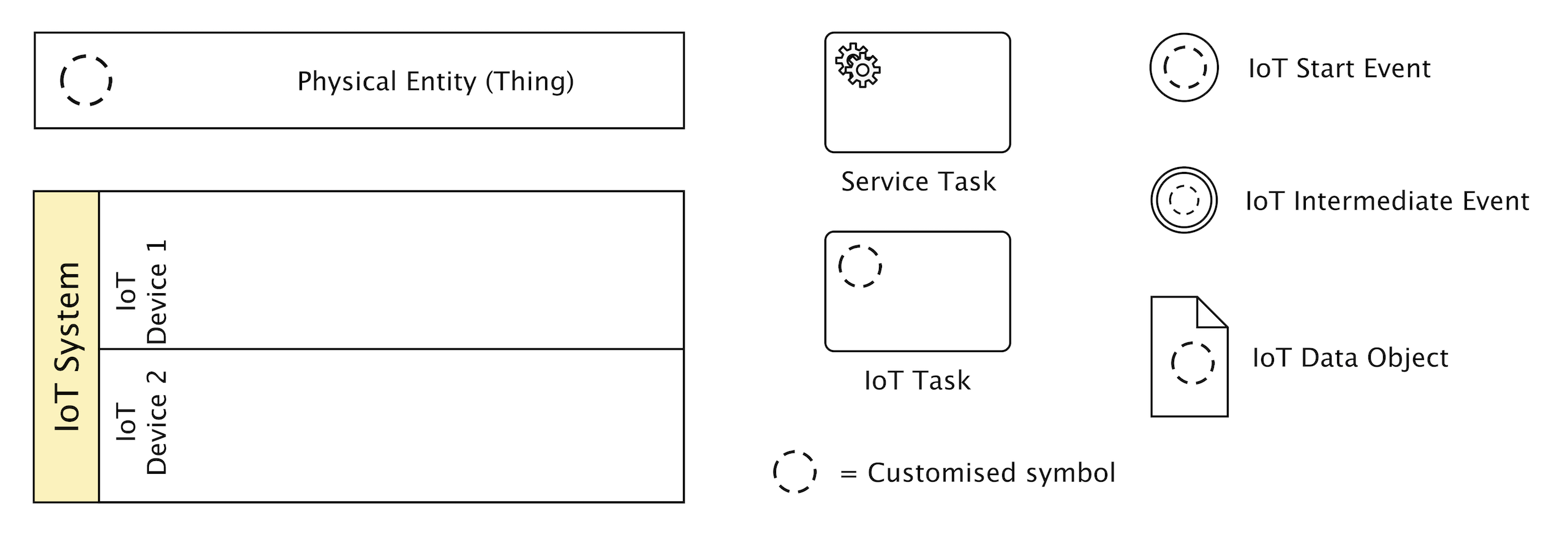}
    \caption{
    		Common notations used in IoT-driven BPMN.
	}
    \label{fig_BPMNElements}
\end{figure}


\medskip
\begin{figure}[h]
  \centering
    \includegraphics[width=0.85\textwidth]{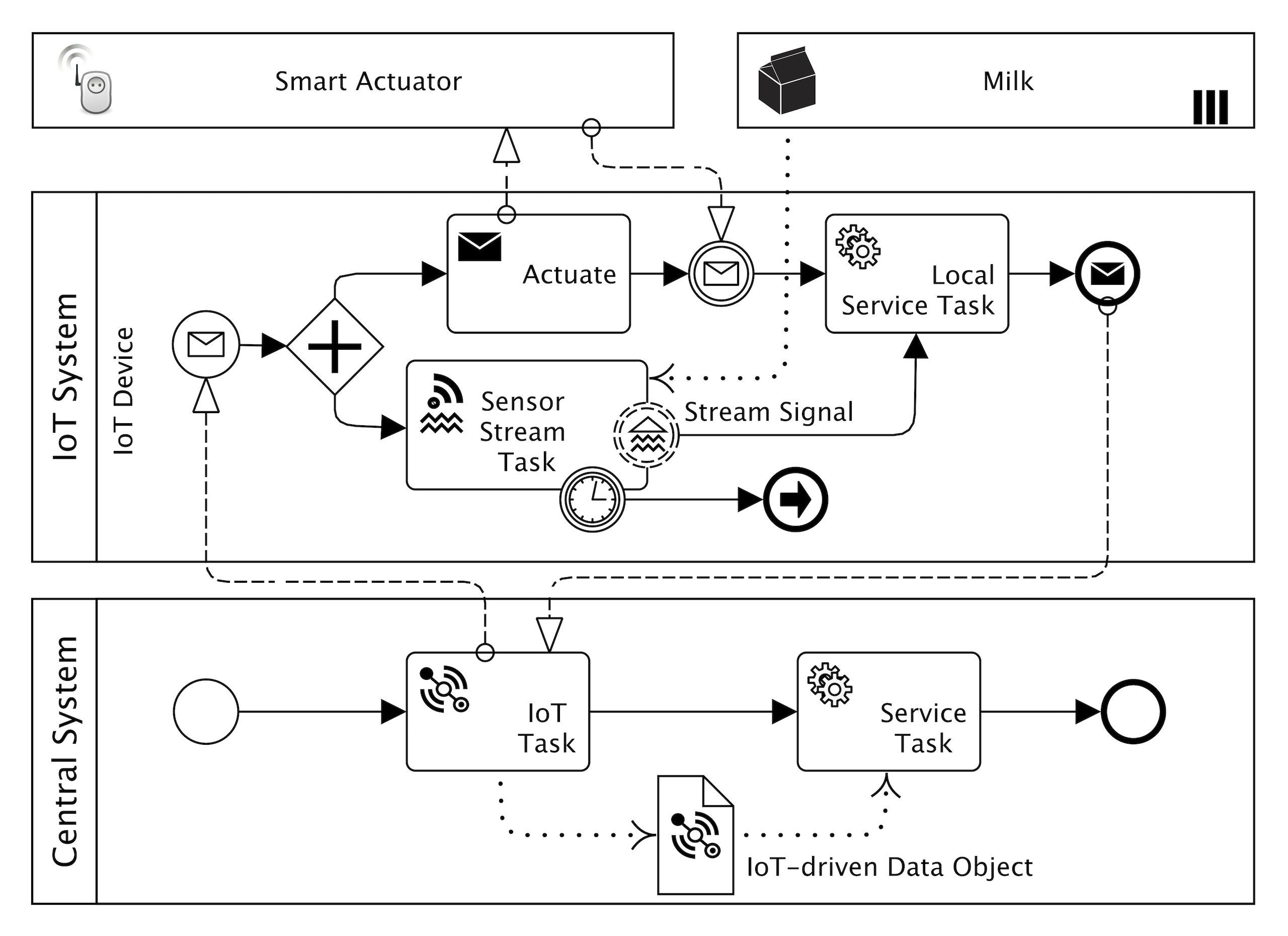}
    \caption{IoT-driven business process workflow.}
    \label{fig_BPMN4IoTExample}
\end{figure}


\noindent 
\textbf{IoT System Process and IoT Device Activity}. 
IoT System Process, or {\it IoT Process}, represents the workflow that involves IoT-related events and activities. In BPMN-based approaches such as IoT-A \cite{Meyer2013}, Sungar et al.~ \cite{Sungur2013} and makeSense \cite{Tranquillini2012}, they notate IoT Process as the Pool (see IoT System in Fig~\ref{fig_BPMNElements}). Specifically, the IoT-A project \cite{Meyer2013} has further modelled IoT device activities in Lane (e.g., IoT Device 1 and 2 in Fig~\ref{fig_BPMNElements}) to separate them from the general processes that excludes the detailed IoT tasks. In contrast, makeSense proposed a different design, which specifically separated the IoT-related processes entirely to different Pool. For example, Fig.~\ref{fig_BPMN4IoTExample} shows an example in which the model separates the IoT-related processes from the Central System.

\medskip
\noindent 
\textbf{Physical Entity}.~Meyer et al.~ \cite{Meyer2015} specifically presents a thorough analysis about how BPMN should signify the real world physical entities such as a chocolate, a bottle of milk, an animal. Consider that  the system may interact with physical entities with heterogeneous protocols,  Text Annotation and Data Object are both less feasible to represent the physical entities. Therefore, their discussion result shows that utilising Participant notation in BPMN is prevailing for  notating physical entities (e.g., Physical Entity (Thing) in Fig.~\ref{fig_BPMNElements}).

\medskip
\noindent 
\textbf{Actuator}.~
An actuator is a controller type device that can perform a certain command to physical objects. For example, an IoT system can remotely switch on or off a light via an interconnected light switch actuator. Commonly, existing frameworks utilise Task or Service Task to notate the actuators involved activities in BPMN. Though, in the work of Yousfi et al \cite{Yousfi2015,Yousfi2016}, they further introduced the specific {\it start event} and the {\it intermediate event} BPMN notations for clarifying what kind of actuator devices are used in the events. 

In general, the approach to introduce IoT-related tasks, events in BPMN is to replace the symbols of notation (e.g. the dashed line circle in Fig.~\ref{fig_BPMNElements}). Note that in this paper, we classify the camera, microphone, tag readers specified in the work of Yousfi et al \cite{Yousfi2015,Yousfi2016} as actuator based on the description by Guinard et al. \cite{Guinard2010}.

\medskip
\noindent 
\textbf{Sensor}.~
IoT systems utilise sensors to acquire specific data such as brightness, temperature, an entity's movement, moving direction and so on. Overall, existing frameworks tend to model sensors differently. The activities of sensors are usually designed as the extension of Task \cite{Meyer2013,Yousfi2015} or the extension of Service Task \cite{Sungur2013} with a specific symbol. Additionally, Yousfi et al. ~ \cite{Yousfi2015} further introduced Sensor Events as the same purpose as they applied to the actuators described in the previous paragraph.


\medskip
\noindent 
\textbf{Event Streaming}~
is a less studied, but an important mechanism in IoT scenarios. Although existing BPMS support single events, they lack feasible integration across the process modelling, the process execution and the IT infrastructure layer. Hence, Appel et al.  \cite{Appel2014} introduced specific element---Stream Process Unit (SPU) to integrate the continuous event streaming processes.

The primary differences between the SPU task and the existing BPMN elements such as Service Task, Loop or parallel operation are: (1) SPU is based on advertisement/subscription of event streaming process that operates in isolation from the main process workflow; (2) SPU performs the process continuously to fulfil a single process, which is different to the Loop that is repeating certain processes (e.g., sensor stream task and stream signal in Fig.~\ref{fig_BPMN4IoTExample}).


\medskip
\noindent 
\textbf{Intermediary Operation.}~
It represents a process that is responsible to perform advanced activities based on the sensory information. Most of existing works \cite{Dar2015,Yousfi2015,Meyer2015} consider the sensory activities as the processes to collect the raw context data. However, Sungur  \cite{Sungur2013} considers that the system should process the raw context data (e.g., interpreted to a meaningful information for the user) before sending it to the requester. Such a requirement is less considered in the other works because it is commonly expected that the application will process the raw context data based on the need of the requester. Besides, the requesters may have their own algorithm to process the data. However, in the future inter-organisational IoT environment, the data collector may also be interested in providing additional context interpreting service to their data consumers. 


\medskip
\noindent 
\textbf{Specific Data Object.}~
Data Object in classic BPMN represents a data or a file that is transmitted from one activity to another. Commonly, it is a one-time transmission. 

In uBPMN \cite{Yousfi2015}, the authors introduced {\it Smart Object} element, which is the sub-class of {\it data object} in BPMN. The {\it Smart Object} element consists of the type attribute to describe the source of the data (e.g., the data was collected by Sensor Task or Reader Task). 

In SPUs \cite{Appel2013,Appel2014}, authors introduced {\it Event Stream} data object, which is different to the classic data object in BPMN that represents one single flow of data transmission. The {\it Event Stream} data object represents the independent streaming data that is being continuously inputted or outputted via the workflow system. Such an approach can specify the sensory information streaming scenario in IoT, which is not considered in classic BP model design.




\medskip
\noindent  
\textbf{Discussion}

\smallskip
\noindent 
Most frameworks introduced IoT elements to adapt to specific scenarios. The most conflicted element defined among the literatures is the sensor element. Some works prefer to model sensory devices as individual information systems, but some works prefer to hide the details and only consider the corresponding sensor tasks. From the perspective of  the real world IoT implementations (e.g. \cite{Chang2015icws}), sensor devices are connected to an IP network (either via mediator, gateway, IPv6 or 6LoWPAN \cite{Shelby2011}), and ideally can be accessed as URI-based RESTful service. Since the standard such as BPMN 2.0 supports URI-based Service Task interaction, it is not clear that differentiating the Sensor Task from  Service Task will bring much effort to the process modelling. On the other hand, process such as event streaming \cite{Appel2014}, which has not yet been considered in classic BP modelling, is necessary to be addressed as a new element.

Commonly, existing modelling approaches have not considered about mobile IoT devices in their model specification. Mobile IoT devices (e.g., wearable devices, flying devices, handheld devices, etc.) have dynamic states, including their location, moving direction, connectivity and so on. These states highly influenced the process operation. For the transparency and agility purposes, this context needs to be considered in the model in order to quickly react to the events. If the assumption is to rely on the cloud middleware to handle the events as separated from the main management system, it loses the transparency purpose.



\subsection{Implement/Configure Phase}
The {\it implement/configure phase} represents how the system transforms the abstract BP model to the machine-readable and machine-executable software program. The challenges in this phase mainly involve in the lack of corresponding tools. Commonly, BP modellers design the BP models in graphical tools, and the tools may generate the machine-readable metadata in order to let the workflow engine to execute the processes. However, the common tools such as Activiti (\url{http://activiti.org}), Camunda (\url{https://camunda.com}), BonitaBPM (\url{http://www.bonitasoft.com}), Apache ODE (\url{http://ode.apache.org}) do not support many of the protocols used in IoT devices. e.g., CoAP and MQTT. Currently, there is no corresponding tool that can address all the protocols used by the IoT devices. A common approach is to introduce middleware layer to leverage the embedded service from IoT devices with the common SOAP or REST Web services. However, without the proper solution, the system may sacrifice the transparency (in BPM perspective), performance (extra overhead) and agility (reaction of run-time events).

\smallskip
\subsubsection{Machine Executable BP Model Approaches}
Currently, from the literature studied in this paper, BPEL, BPMN and XPDL are three common standard technologies used in implementation.


\medskip
\noindent \textbf{BPEL for IoT.}~
Web Services Business Process Execution Language (WS-BPEL; or BPEL in short) is an executable and an interoperable service composition language in which BPEL is fully WS-* compliant. Ideally, the BPEL metadata is executable in any standard compliant engines as long as the engines have fetched the required corresponding descriptions (WSDL, XML schema etc.).~
Notably,  both academic research projects \cite{Loke2003,Glombitza2011,Yang2012,Wu2012,Chang2015scc} and industrial solutions such as WSO2 IoT Server \cite{WSO2IoT}, SiteWhere (\url{http://www.sitewhere.org/}) and Oracle SOA Suite \cite{OracleSOAIoT} have introduced BPEL-based BPMS4IoT.~
However, the standard itself does not natively support RESTful service invocation, which is now the primary approach to providing services from IoT devices. According to the W3C standard, WSDL 2.0 can describe HTTP method-based invocation. Unfortunately, WS-BPEL does not support WSDL 2.0. On the other words, current BPEL standard is mainly for WSDL 1.0 and SOAP-based service composition only. Although numerous commercial BPEL engines have support for RESTful HTTP service, it is not sufficient for the other common protocols used in IoT devices. Moreover, BPEL supports  orchestration only. It does not support the choreography natively.



\medskip
\noindent \textbf{BPMN for IoT.}~
Business Process Model and Notation (BPMN) has been the most popular approach to the literature studied in this paper. These works include IoT-A~\cite{Meyer2013},~
VITAL \cite{VITAL}, Canraca\c{s} et al. \cite{Caracas2011}, BPMN4WSN  \cite{Sungur2013}, MOPAL \cite{Peng2014}, SPUs \cite{Appel2014}, makeSense~\cite{Casati2012,Tranquillini2012,Tranquillini2015}, uBPMN \cite{Yousfi2015}, Dar et al. \cite{Dar2015}. Although originally, BPMN is only a  graphical BP model standard introduced by the Object Management Group (OMG), OMG also introduced the metadata form in XML format. BPMN provides a flexibility for designers to introduce their own extension of BPMN elements. Hence, it is possible to introduce new IoT entities and elements semantically using BPMN, XML schema and semantic description methods. Although the extension grants flexibility, there is no interoperability between different design models because they are tightly bounded onto the corresponding execution engine and hence result in isolated solutions. In the other words, the BPMS designed based on makeSense cannot cooperate with uBPMN-based BPMS. 


\medskip
\noindent \textbf{XPDL for IoT.}~
XML Process Definition Language (XPDL; \url{http://www.xpdl.org/}) is a standard introduced by the Workflow Management Coalition (\url{http://www.wfmc.org/}) for interchanging the graphical business process workflow models to XML-based meta-models. Numerous projects have used XPDL to enable workflow execution on mobile devices \cite{Chou2009,Chen2011} or other CPS devices \cite{Kefalakis2011}. Ideally, XPDL is a continuously updated standard to be compatible with the latest BPMN standard. Although its primary purpose is to serve as an XML interchange of BPMN, developers can also use it as the meta-model of the other BP workflow definitions. Accordingly, the standard's document indicates that XPDL is a standard for interchanging any type of workflow model to machine readable code. Hence, compared to BPEL, XPDL can be more flexible in terms of implementation and execution.

\medskip
\noindent \textbf{Discussion}

\smallskip
\noindent
Overall, existing BP model standards natively do not support the need for the distributed process in the near future IoT systems. Among them all, BPMN and XPDL provide the flexibility of the extension, hence, they are applicable when interoperability is not the concern of the system. Although some other BP modelling approaches exist (e.g. Petri-Net model based system \cite{IgeiKaneshiro2014,Thacker2010}), there is no corresponding linkage between the model and the executable program.

\smallskip
\subsubsection{Process Execution Approaches}\label{sec:BPE}

The {\it process execution} represents how the workflow engines execute the transformed machine-readable process model. In general, 
the workflow engines are hosted either on the central management server or on the participated IoT devices in the edge network. 

Considering the classical centralised system, where the system does not require performing complex processes on participants, the workflow engines only hosted in the central system, in which  the IoT devices can operate as the regular request/response operation-based services. Conversely, if the system requires process distribution, the IoT devices may need to embed process execution engines. Overall, there are three types of situations in BPMS4IoT. We described them below.

\begin{figure*}[t]
    \centering
    \begin{subfigure}[t]{0.27\textwidth}
        \includegraphics[width=\textwidth]{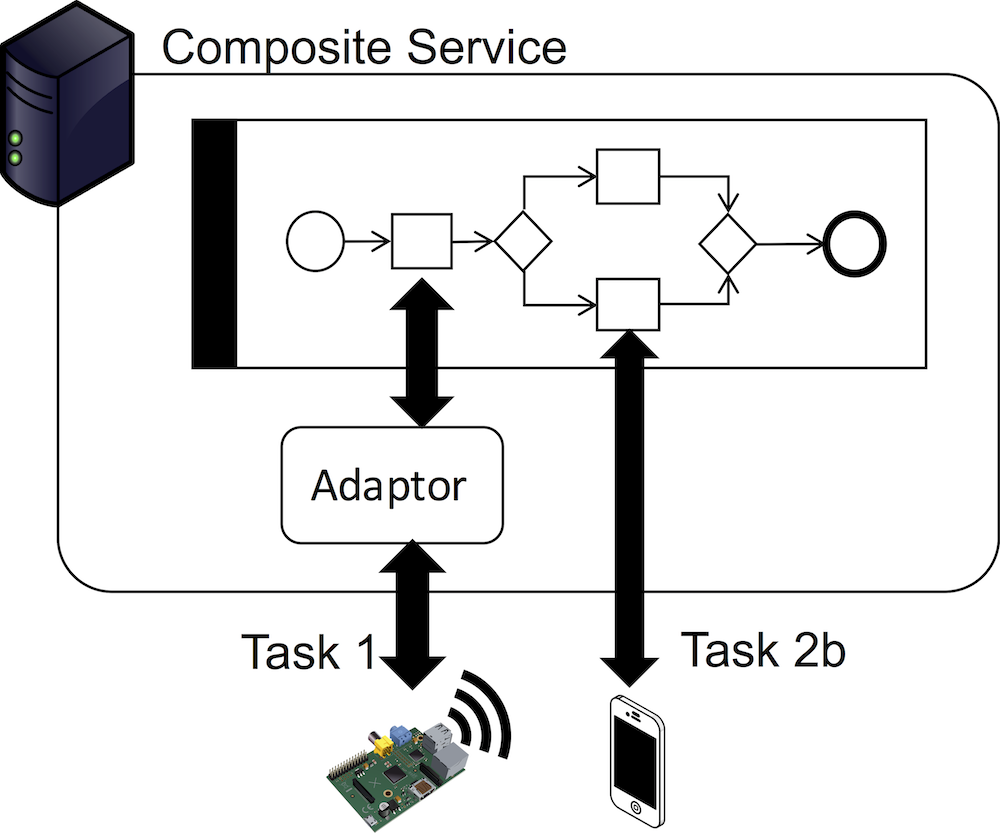}
        \caption{Device Participating in BP}
        \label{fig:deviceInBP}
    \end{subfigure}
    \hfill%
    \begin{subfigure}[t]{0.32\textwidth}
        \includegraphics[width=\textwidth]{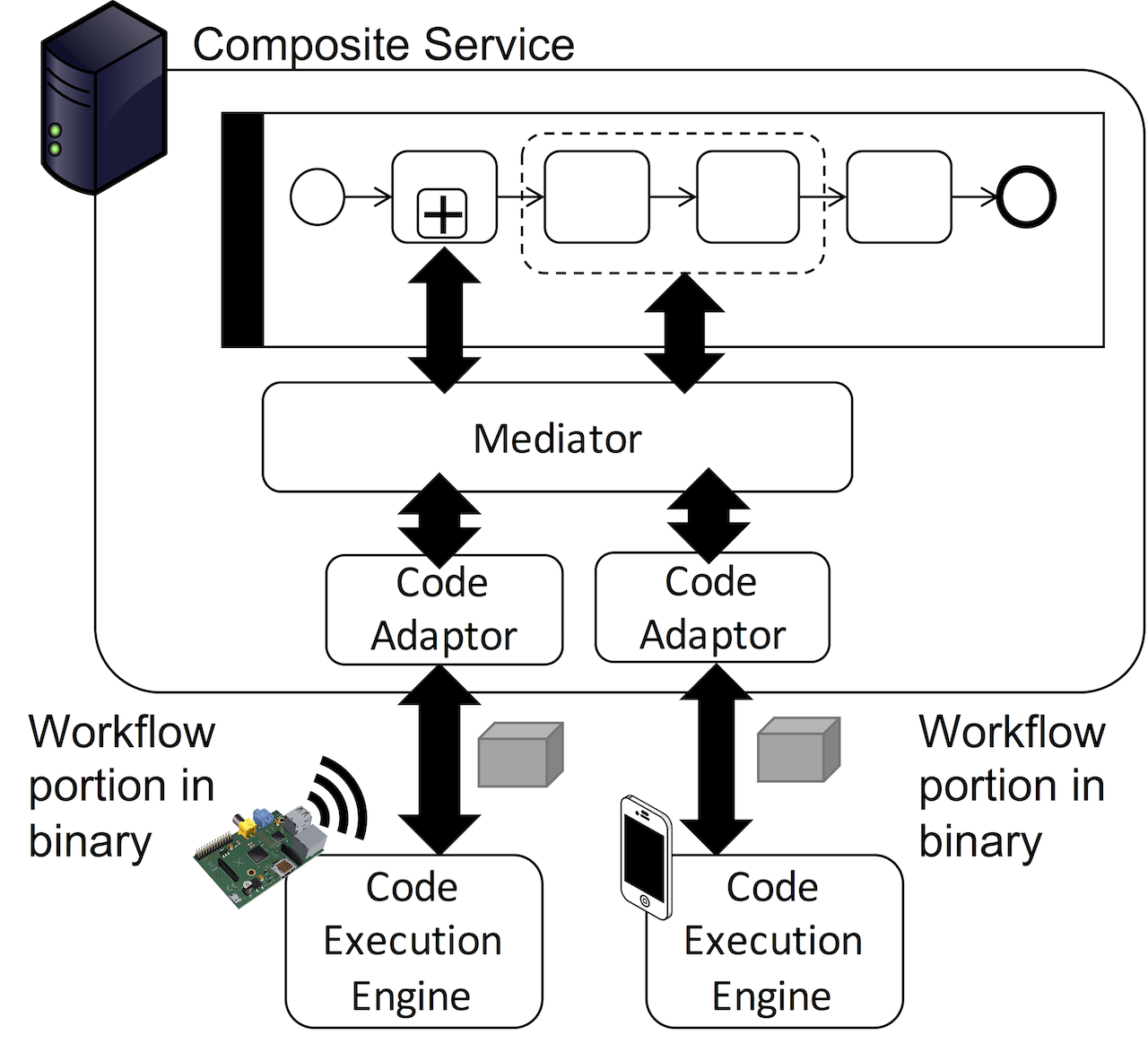}
        \caption{Devices execute code-converted BP workflow}
        \label{fig:deviceDoCode}
    \end{subfigure}
    \hfill%
    \begin{subfigure}[t]{0.37\textwidth}
        \includegraphics[width=\textwidth]{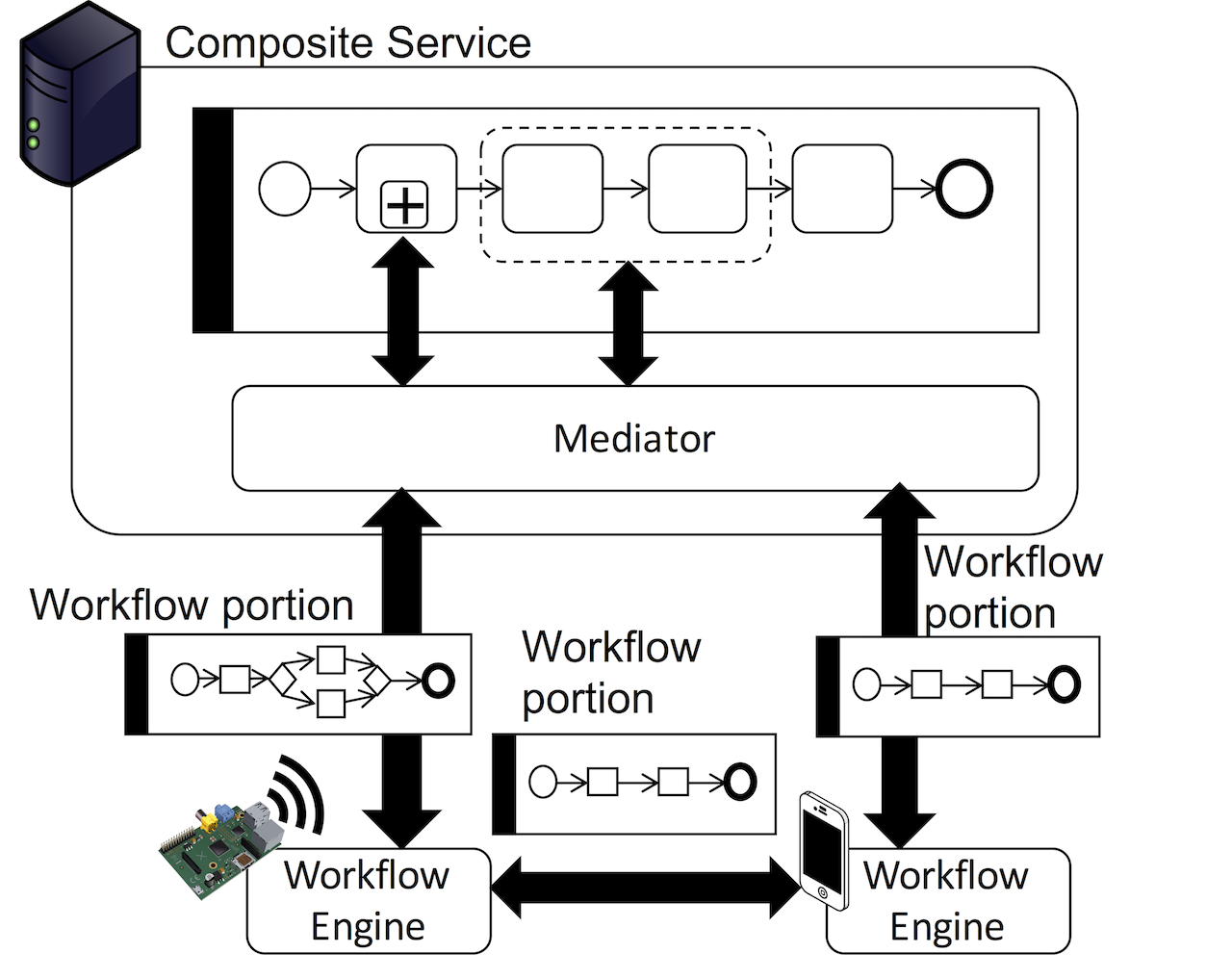}
        \caption{Devices execute BP workflow}
        \label{fig:deviceDoBP}
    \end{subfigure}
    \caption{Different approaches for integrating physical things in BPMS}\label{fig:animals}
\end{figure*}

\medskip
\noindent \textbf{Participating in BP.}~
In this case, the participative node only performs certain tasks based on request/response or publish/subscribe mechanism (see Fig.~\ref{fig:deviceInBP}). For example, in the nursing home scenario described in \cite{Pryss2015}, the hospital's BPMS can remotely assign tasks to the front-end nurses via their mobile devices. Moreover, since the mobile devices act as the mediators, they can either allocate manual tasks to the nurses or retrieve the data from the patients to the remote hospital BPMS. 

Practically, the front-end devices used in this approach can simply enable socket channels to maintain the communication between themselves and the central system. Alternatively, the front-end devices can embed Web services as the service providers \cite{Srirama2006,Liyanage2015} (e.g. Task 2b in Figure~\ref{fig:deviceInBP}), which provide services directly without maintaining a long period communication channel with the distant central system.

%

\medskip
\noindent \textbf{Executing BP Model-Compiled Code.}~
In this approach, the system translates the BP model metadata generated from the BP model editor to the executable binary code or to a specific programming language (see Fig.~\ref{fig:deviceDoCode}). In particular, this approach may be more suitable for low-level resource constrained devices to execute the BP models. Generally, in this approach,  the system relies on the middleware technologies to translate the BP model to the executable code, then send the code to the IoT devices for execution. Particularly, numerous BPMS4IoT frameworks \cite{Glombitza2011,Caracas2011,Casati2012,Tranquillini2012,Tranquillini2015} have utilised this approach to enable IoT/WSN devices participating in BP execution without the need of embedding complex software middleware components (e.g., workflow engine) on the devices.

\medskip
\noindent \textbf{Executing Standard BP Model.}~
Directly porting or implementing a workflow engine on the IoT device enables the best flexibility (see Figure~\ref{fig:deviceDoBP}). Moreover, it also enables a flexible way of performing process choreography between the back-end cloud service and the front-end IoT devices or between numerous front-end IoT devices, in which the front-end IoT devices situated in edge network are executing the workflow \cite{Pryss2015}. However, such a mechanism requires higher computational power devices (e.g., high-end smartphones). In the meantime, numerous research projects have introduced the standard-based workflow execution engines for mobile operating systems. 

\medskip
Table~\ref{mobileworkflowengines} lists a number of featured workflow engines designed for mobile devices. Further, the table also describes certain additional elements supported by the workflow engines.
\begin{table*}[t]
\footnotesize
\centering
\caption{Comparison of workflow engines for mobile devices}
\label{mobileworkflowengines}
\renewcommand{\arraystretch}{1.2}
\begin{tabular}{l||l|l|l|l|l}
\hline
                        & \textbf{Standard} & \textbf{\begin{tabular}[c]{@{}l@{}}Workflow\\ Distribution\end{tabular}} & \textbf{\begin{tabular}[c]{@{}l@{}}Protocol\end{tabular}}   & \textbf{Awareness}                                           & \textbf{Platform}                                    
\\ \hline \hline
\pbox{20cm}{\textbf{Sliver}\\\cite{Hackmann2006}}         
& BPEL              & ---                                                                      & SOAP                                                                    & ---                                                          & Java ME                                              
\\ \hdashline[1pt/2pt]
\pbox{20cm}{\textbf{CiAN}\cite{Sen2008}}          
& BPEL              & \begin{tabular}[c]{@{}l@{}}Device-to-\\ Device\end{tabular}              & SOAP                                                                    & ---                                                          & Java ME                                              
\\ \hdashline[1pt/2pt]
\pbox{20cm}{\textbf{Pajunen et al.}\\\cite{Pajunen2007}}
& BPEL              & ---                                                                      & SOAP                                                                    & ---                                                          & Java ME                                              
\\ \hdashline[1pt/2pt]
\pbox{20cm}{\textbf{AMSNP}\\\cite{Chang2012icsoc}}      
& BPEL              
& \begin{tabular}[c]{@{}l@{}}Cloud and \\ device\end{tabular}              & \begin{tabular}[c]{@{}l@{}}RESTful\\ HTTP\end{tabular}                  & \begin{tabular}[c]{@{}l@{}}Resource-\\ aware\end{tabular}    & iOS                                                  
\\ \hdashline[1pt/2pt]
\pbox{20cm}{\textbf{SPiCa}\\\cite{Chang2014mum}}       
& BPEL              
& \begin{tabular}[c]{@{}l@{}}Device-to-\\ Device\end{tabular}              & \begin{tabular}[c]{@{}l@{}}RESTful\\ HTTP\end{tabular}                  & \begin{tabular}[c]{@{}l@{}}Resource-\\ aware\end{tabular}    & iOS                                                  
\\ \hdashline[1pt/2pt]
\pbox{20cm}{\textbf{SCORPII}\\\cite{Chang2015scc}}      
& BPEL              
& \begin{tabular}[c]{@{}l@{}}Cloud and\\ device\end{tabular}               & \begin{tabular}[c]{@{}l@{}}RESTful\\ HTTP\end{tabular}                  & \begin{tabular}[c]{@{}l@{}}Resource-\\ aware\end{tabular}    & Android                                              
\\ \hdashline[1pt/2pt]
\pbox{20cm}{\textbf{MAPPLE}\\\cite{Pryss2011}}     
& Cutomised         
& \begin{tabular}[c]{@{}l@{}}Cloud and\\ device\end{tabular}               & \pbox{20cm}{Stand-\\alone}                                          
& \begin{tabular}[c]{@{}l@{}}Resource-\\ aware\end{tabular}    & .NET                                                 
\\ \hdashline[1pt/2pt]
\pbox{20cm}{\textbf{EMWF}\\\cite{Chou2009}}           
& XPDL              
& ---                                                                      & ---                                                                     & ---                                                          & \begin{tabular}[c]{@{}l@{}}Windows\\ CE\end{tabular} 
\\ \hdashline[1pt/2pt]
\pbox{20cm}{\textbf{ERWF}\\\cite{Chen2011}}        
& XPDL              
& ---                                                                      & ---                                                                     & \begin{tabular}[c]{@{}l@{}}Environment-\\aware\end{tabular} & SISARL                                               
\\ \hdashline[1pt/2pt]
\pbox{20cm}{\textbf{Dar et al.}\\\cite{Dar2015}}
& BPMN              
& \begin{tabular}[c]{@{}l@{}}Cloud and\\ device\end{tabular}               & \begin{tabular}[c]{@{}l@{}}CoAP/\\MQTT/\\RESTful\\HTTP\end{tabular} 
& ---                                                   
& Android                                              
\\ \hline
\end{tabular}
\end{table*}
\begin{itemize}
\renewcommand\labelitemi{$\bullet$\;}
\item {\it Standard}---describes which workflow modelling or description standard the engine supports. Overall, only Presto \cite{Giner2010} and Dar et al. \cite{Dar2015} support BPMN. Fundamentally, BPMN is a graphical tool, it requires an extra mechanism to convert the graphical BP model to machine-readable code. Hence, it is understandable that most engines have chosen to support XML-based modelling standard (i.e. BPEL and XPDL).
\smallskip
\item {\it Workflow distribution}---illustrates how does the engine handle the workflow migration between different entities. Specifically, handling workflow migration between different entities (e.g. between cloud and mobile or between mobile and mobile) is an important mechanism to support the machine-to-machine communication in the near future IoT applications \cite{Dar2015}. Hence, it is foreseeable that such a feature may become a requirement in future BPMS4IoT.
\smallskip
\item {\it Protocol}---describes the supported protocols of the engine. Generally, earlier engines \cite{Hackmann2006,Sen2008,Pajunen2007} aim to be fully compliant with the Web service standard. Hence, SOAP was their primary consideration. However, since the wireless IoT devices are usually resource constraint, recent approaches intend to apply lightweight protocols. Although the workflow description standards (e.g. BPEL and BPMN) natively do not include the definition of lightweight protocol-based service invocation such as HTTP-based RESTful Web services, Constrained Application Protocol (CoAP) or even the customised lightweight protocols   \cite{Pryss2011,Giner2010}, many engines have supported the need.
\smallskip
\item {\it Awareness}---denotes whether the engine supports a certain awareness mechanism to handle the runtime events or not. Overall, a number of existing engines have supported {\it resource-awareness}, which represents the strategy to improve the reliability of the software or hardware resource used in executing the workflow tasks. Specifically, ERWF \cite{Chen2011} supports {\it environmental-awareness}, in which the workflow engine can react to the environmental changes at runtime to maintain the performance.
\smallskip
\item {\it Platform}---describes which platform or operating system the engine supports. As the table shows, most projects developed their engines based on smartphone OS, except ERWF \cite{Chen2011}, which is targeted on the SISARL sensor devices (\url{http://www.sisarl.org/}) that have lower power than regular smartphones.
\end{itemize}

\medskip
\noindent \textbf{Discussion}

\smallskip
\noindent
The drawback of the {\it executing BP model-compiled code} approach is that every time the process model changes, the system needs to  again convert the model to the executable source code and to deploy  the code to the IoT devices. Conversely, the embedded workflow  engine approach can perform rapid changes in the model and execution. However, it may also consume more hardware resources of the device.

The engines described in Table~\ref{mobileworkflowengines} brings the possibility of distributed business process workflow execution between {\it cloud-to-mobile}, and {\it mobile-to-mobile}, in which the system can support choreography-based composition towards bringing the highly flexible and scalable mobile cloud-based BPMS,  which can be useful for many IoT systems such as the logistics and the AAL use cases described previously in Section~\ref{sec:usecases}.

\subsection{Run and Adjust Phase}

The {\it run and adjust phase} represents the system execution runtime after the deployment of the BPMS. In general, this phase does not involve any redesign and neither the new implementation. Specifically, since the run and adjust phase only performs the predefined management activities \cite{Aalst2013}, the BPMS should provide adaptive mechanisms for the system administration. For example, the system should record the runtime execution history, handle the events and allow the adjustments. 

Following are the major mechanisms the BPMS4IoT should address in the {\it run and adjust phase}.

\subsubsection{Monitoring and Control}
BPMS4IoT involves various front-end devices formed edge networks in which groups of static or mobile devices are participating (e.g. collaborative sensing). Commonly, a design such as a cloud service-based IoT environment \cite{Botta2016}, the system requires multiple layers to enable the activity monitoring of the edge networks. In other words, the IoT devices need to connect to the broker, sink or super-peer (the head of an edge network) devices in order to enable the communication between themselves and the back-end systems using publish-subscribe protocols (e.g. MQTT).

{\it Monitoring} in {\it run and adjust phase} involves two types: {\it process monitoring} \cite{Aalst2013} and {\it device status monitoring} \cite{Dar2015}. 
\begin{itemize}
\item {\it Process monitoring} is the comprehensive system monitoring, which involves how the system operates the IoT activities and how it handles the runtime events. At this stage, the system collects the  {\it process monitoring} records log files for further analysis in order to improve the system in the {\it redesign phase}. 
\item {\it Device status monitoring}. Since the activities of  BPMS4IoT involve a large number of wireless network devices, the runtime device failure can sometimes cause the entire system to fail. Hence, {\it device status monitoring} also plays an important role in BPMS4IOT. 
\end{itemize}

Monitoring edge network is a crucial task because it can involve many factors such as the hardware damage occurs in the edge network device, the cluster head of the edge network lost its connectivity with the other edge peers, the data broker node lost its connectivity to the backend system due to either the hardware or the Internet provider failure. Overall, most existing BPMS4IoT frameworks have not broadly studied about the runtime device failure issues. Among the existing frameworks, \cite{Dar2015} applied a generic approach to handle the runtime device failure. In their approach, the system assigns a periodical report task to each front-end device. If the central server does not receive the report from a particular device exceeding the timeout threshold, the system will consider the device is failed. Hence, the system will perform the substitution.

\subsubsection{Fault Tolerant}
IoT paradigm involves a large number of wireless networks-connected devices. Generally, these devices have less stability in connection and with the low battery capability. Therefore, a proper system needs to consider the resource-constrained devices and to support the corresponding solutions, such as reactively replacing the failed devices/activities with substitutions immediately and autonomously in order to retain the processes \cite{Dar2015}. Further, in order to optimise the system to react or even prevent the failure, the system needs to distribute and govern a certain business processes to the edge network. For example, considering the unreliable mobile Internet connection, the authors in \cite{Peng2014} has proposed a framework for distributing and executing tasks at the edge network mobile nodes in offline mode. Explicitly, with such a mechanism, when an edge network lost its Internet connection with the distant management system, the devices in the edge network can still continue the processes. Furthermore, they can send the process output and the monitored records to the distant management system once they are back in connection.
%

\subsubsection{Context-awareness}\label{sec:context-awareness}
\textit{Context is any information that can be used to characterise the situation of an entity. An entity is a person, place, or object that is considered relevant to the interaction between a user and an application, including the user and applications themselves}
\cite{Dey2001}. BPMS4IoT needs to address context-awareness in terms of contingencies, personalisation, efficiency \cite{Sheng2014}.

\smallskip
\noindent
\textbf{Contingencies} involve the unpredictable connectivity and accessibility of pervasive services and wireless network devices. Generally, there are two schemes to address contingencies in BPMS4IoT: {\it proactive} and {\it reactive}. 
\begin{itemize}
\item {\it Proactive} scheme aims to prevent the occurrence of problems from IoT devices at runtime. For example, in the MOPAL project \cite{Peng2014}, authors have defined a certain context-aware rules (e.g. current CPU capability, battery level, geographical location etc.) that constrain the workflow task execution on the IoT devices. 
\item {\it Reactive} scheme commonly seeks for substitution for the workflow task execution. In the SOA-based BPMS for home automation system proposed by Chang and Ling \cite{Chang2008}, the connected devices belong to specific categories depending on their associated context. For example, both workflow tasks---``sound alarm" and ``switch on TV$\rightarrow$raise TV volume"---can generate same type context---``loud noise" to wake up the user. Hence, when the default setting---``sound alarm" is failing due to any reason, the system can trigger the substitution, which is ``switch on TV$\rightarrow$raise TV volume" to achieve the same purpose.
\end{itemize}

\smallskip
\noindent 
\textbf{Personalisation} involves service provisioning based on the requesters' preferences. For example, in the smart ubiquitous computing domain \cite{Peng2014,Dar2015}, context- awareness has usually considered the entity's context. Ordinarily, the entity in most cases is the human users themselves, and the context can be the person's heartbeat rate, blood, breath, physical movement, etc. Comparatively, works in the AAL domain \cite{Loke2003,Yousfi2015,Chang2015scc} may consider environmental context, such as the temperature, the noise level, the density of the crowd, in which the central entity (the user) does not have direct control of them.

\smallskip
\noindent
\textbf{Efficiency}~
involves energy efficiency, the cost of deployment and cost of communication.

\begin{itemize}
\item {\it Energy efficiency}---is one of the major concerns in BPMS4IoT. In order to conserve the energy of IoT devices, Caraca{\c{s}} et al. \cite{Caracas2011,Caracas2012thesis}  specified the {\it time} context factor in the BP model. Basically, in their BP model, each IoT task should associate with Timers for controlling the sleep/wake up time of the device. Alternatively, inter-organisational collaborative data brokering strategy is also a promising approach. For example, in the work of Chang et al. \cite{Chang2015icws}, the IoT devices, which belong to different organisations, are collaboratively brokering the data to their distant backend server. In general, the data collector device will automatically seek for collaboration when its battery is getting low. Ideally, such an approach can reduce the unnecessary energy consumption compared to sending data individually.
\smallskip
\item {\it Deployment}. Deploying IoT devices individually can be costly. One possible strategy is to re-use the already-deployed system and provide an integration platform such as SOA-based cloud service that can enable the need and also provide the interoperability between the BPMS of different organisations. A BPMS4IoT can compare between the cost of deploying its own devices and the cost of utilising the third party's resources. In contrast, the second option can result better efficiency in deployment. For example, in GaaS project  \cite{Wu2012}, the cloud-based gateway platform enables IoT systems from different parties to work together. Similarly, in Adventure project \cite{Schulte2014}, the system utilises the virtual machine-based cloud to provide IoT-based manufacturing ERP instance, which facilitates the ERP procedure and also help manufacturers to find partners for the production.
\smallskip
\item {\it Communication}. The cost of communication at runtime can be very dynamic, not only due to the traffic changes of the Internet Service Provider (ISP)-side but also influenced by the number of IoT devices involved and where the requester is located. For example, when a ubiquitous IoT application requires using WSN in its surrounding, the process is mainly relying on the distant data centre. Explicitly, such an approach  can cause high latency when the Internet connection is not in good condition. Therefore, many approaches utilise proximity-based resources for data acquisition and processing intensive applications. For example, the concept of mobile ad-hoc cloud computing is possible to cater such need \cite{Loke2015}. Alternatively, cloudlet-based MCC \cite{Gao2012} is also an efficient option. Further, the extension of cloudlet known as Fog computing \cite{Bonomi2012} is getting industry attention recently. Although existing BPMS4IoT frameworks have not yet explicitly address the Fog computing-enabled system.
\end{itemize}

\subsubsection{Scalability}\label{sec:scalability}
Scalability is one of the major requirements for management of large scale connected devices \cite{Conti2012,Teixeira2011,Issarny2011,Borgia2014}. In existing BPMS4IoT frameworks, scalability involves two specific topics: the growing volume of stream data and the growing number of BP participative devices.
\begin{itemize}
\renewcommand\labelitemi{$\bullet$\;}
\item The growing volume of stream data derived from various data sources cross different parties. The large volume of data is utilised to provide the need of domain specific applications such as disaster recovery, urban computing \cite{Salim2015}, social computing \cite{Wang2007,Chang2012} etc. In order to handle the large volume of stream data from different sources, a common approach is to utilise service-oriented Enterprise Service Bus (ESB) architecture together with elastic cloud computing resources \cite{Appel2014}.
\item The growing number of BP participative devices represents the execution of BP will involve numerous heterogeneous IoT devices. Existing works, which have addressed this topic, can be classified into two types: centralised and decentralised.
	\begin{itemize}
	\item In centralised solution, Wu et al. \cite{Wu2012} introduced the Gateway as a Service (GaaS)-based architecture that enables inter-organisational collaboration among the IoT devices. The GaaS architecture is based on the foundation of cloud services (Infrastructure as a Service, Platform as a Service, Software as a Service). By integrating the cloud services and the mediating technology, enterprises can connect their IoT entities as Web of Things (WoT) collaboratively.
	\item In decentralised solution, Dar et al. \cite{Dar2015} proposed a framework that enables self-management in edge networks. It is realised by hosting embedded workflow execution engines on the participative IoT devices. Since the workflow execution engines can execute the standard BPMN workflow model, the system can dynamically distribute the processes among the IoT devices in the edge network without pre-establishing the topology. Hence, it achieves the choreography-based BP scalability.
	\end{itemize}
\end{itemize}

\medskip
\noindent \textbf{Discussion}

\smallskip
\noindent
The major purpose of BPM is to optimise the processes of organisations. The BPMS designed for IoT needs to clearly address the challenges in the field. Currently, existing works in BPMS4IoT are still in early stages. Most works are focusing on proposing the solutions for the previous two phases---{\it (re)design} and {\it implement/configure}. Although some works discussed in this section have addressed a few issues in {\it run and adjust phase}, they have not proposed concrete, generic solutions for the specific challenges involved in IoT environment. 

The subjects described in this section highly influence the success of the IoT system when mobile is involved (e.g. in AAL, logistics use cases). Specifically, each of the subjects faces the same challenges as in many MCC solutions. For example, in the literature study proposed by  Fernando et al. \cite{Fernando2013}, the authors have summarised numerous projects that have addressed fault tolerant, context-awareness and scalability in mobile and wireless networks. For this reason, it indicates that many runtime management solutions proposed for MCC can be applied in BPMS4IoT.


\section{Comparison of BPMS4IoT Frameworks}\label{section_comparison}
\begin{figure*}[h]
\footnotesize
\centering
\begin{forest}
  for tree={
    edge path={
      \noexpand\path[\forestoption{edge}](!u.parent anchor) -- +(5pt,0) |- (.child anchor)\forestoption{edge label};},
    grow=0,
    reversed, 
    parent anchor=east,
    child anchor=west, 
    anchor=west,
    if n children=0{tier=word}{}
}
[BPMS4IoT Frameworks,
	[Theoretical IoT-driven Business Process Modelling]
	[Comprehensive]
	[Practical System Integration]
]
\end{forest}
\\
\bigskip
\caption{Taxonomy of BPMS4IoT Frameworks}
\label{fig:taxonomyBPMS4IoTFrameworks}
\end{figure*}
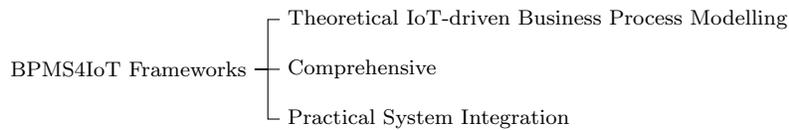

Research projects in BPMS4IoT propose their frameworks for different objectives. Overall, we can classify these frameworks into three types  ((see Fig.~\ref{fig:taxonomyBPMS4IoTFrameworks}): (1) frameworks for theoretical IoT-driven business process modelling; (2) frameworks for the comprehensive solution that cover both modelling and practical system integration; and (3) frameworks for practical system integration only.

In the following discussion of BPMS4IoT frameworks, we divide them into two framework comparison sections: theoretical modelling and practical system integration. Afterwards, we discuss  the feasibility of the frameworks in the different IoT deployment models.

\subsection{Modelling and Architecture Design}
In this section, we discuss the modelling approaches of these frameworks, including the frameworks that focused only on modelling, and frameworks that provided comprehensive solutions. Firstly, we summarise each involved frameworks.

\medskip
\noindent 
\textit{IoT-driven Business Process Modelling Frameworks}
\begin{itemize}
\item \textbf{IoT-A} \cite{Castellani2012,Meyer2011,Meyer2013,Sperner2011} is one of the Seventh Framework Programme for Research and Technological Development (FP7) projects that focus on designing IoT-driven BP models and system architecture for future ERP systems. In the system architecture of IoT-A, the {\it things} of IoT are specifically representing non-electronic physical entities such as chocolate, bottles of milk, animals and so on. Further, {\it things} are “connected” via IoT devices (e.g. RFID reader, wireless sensors, etc.) to information systems as resources. The system can further adapt the resources to atomic or composite IoT services for external applications to interact. Previously, Section~\ref{section_design} has discussed the BP model design of IoT-A. Accordingly, the analysis and the proposed meta-model of BPMN extension in IoT-A project  provide guidance for the (re)design phase of BPMS, especially in clarifying the differences between the real world physical entities, IoT devices and IoT services, which can be a useful foundation for developing BPMS4IoT.
\smallskip
\item \textbf{Sungur et al.}~
\cite{Sungur2013}~
proposed a framework for extending BPMN with WSN elements. They identified the model requirements of WSN based on the characteristics of WSN devices and services. Further, they propose a design for the WSN-driven BPMN extension, which includes a number of WSN-specific notations such as Sense Task, Actuate Task and Intermediary Operation Tasks. Moreover, the framework provides the corresponding XML Schema for each proposed element, and a modelling tool based on the extension of the Web-based Oryx BP editor (\url{http://bpt.hpi.uni-potsdam.de/Oryx}).
\smallskip
\item \textbf{uBPMN} \cite{Yousfi2015,Yousfi2016} is a project aimed to introduce ubiquitous elements in BPMN. The authors defined BPMN Task extension for Sensor, Reader, Collector, Camera and Microphone. Each element also has a corresponding BPMN-Event symbol for Start Event and Intermediate Event. Additionally, uBPMN also introduced an IoT-driven Data Object called Smart Object to represent the data transmitted from the IoT devices. Further, they introduced Context Object to describe the  context attributes (e.g. temperature$\leq$15, container$=$`closed') of a BPMN Flow Element. In general, the framework aims to serve as a guidance similar to the IoT-A project.
\end{itemize}

\noindent
\textit{Comprehensive Frameworks}
\begin{itemize}
\item \textbf{makeSense} \cite{Casati2012,Tranquillini2012,Tranquillini2015}. 
The FP7 project---makeSense aims to provide a comprehensive solution that facilitates the programming and system integration with WSN. As a comprehensive solution, makeSense involves both theoretical IoT/WSN-driven BP model design and practical system integration software. 
\\
In the model design, makeSense introduced two separated BPMN concepts: Intra-WSN Pool and WSN-aware Pool.
\smallskip
\begin{itemize}
\item[$\circ$\;] {\it WSN-aware Pool}---represents regular BPMN Pool, which describes the main workflow of the operation. The WSN-aware Pool associates with Intra-WSN by denoting the IoT/WSN activities in a simplified WSN activity notation. 
\smallskip
\item[$\circ$\;] {\it Intra-WSN Pool}---describes the detailed workflow description of the IoT/WSN tasks. Previously, Fig~\ref{fig_BPMN4IoTExample} has shown a similar design example. 
\end{itemize}
\smallskip
The decomposed design introduced in the makeSense project helps modellers with a clear perspective of the integration system. Further, the project has implemented the proposed modelling approach as an extension of Signavio Core Components called BPMN4WSN editor.
\smallskip
\item \textbf{ASPIRE} \cite{Kefalakis2011}. The modern smart manufactory ERP system utilises RFID technologies to facilitate the supply chain processes. However, such systems commonly require the engagement of low-level RFID-driven programming tasks, which is time-consuming, mainly because of the lack of the standard integration model. In order to overcome the problem, the FP7 project---ASPIRE aims to tackle such issue by introducing a new specification and its practical platform.
\\
ASPIRE project introduces AspireRFID Process Description Language (APDL) specification, which is the extension of XPDL, to leverage EPCGlobal specification with BPMS. Fundamentally, APDL extends the feature of EPCGlobal architecture in generating automatic business process events that can assist the filtering of RFID stream data. 
\\
Accordingly, APDL contains two main concepts for describing the business processes: (1) Open Loop Composite business process (OLCBProc), which describes the BP execution among different individual systems (i.e. the inter-organisational BPMS); (2) Close Loop Composite Business Process (CLCBProc), which describes the execution within one individual system (i.e. the intra-organisational BPMS). 
\\
The project has developed a modelling tool called Business Process Workflow Management Editor (BPWME), which is an Eclipse IDE plugin that allows the modeller to configure the RFID-driven BP for both OLCBProc and CLCBProc level.
\smallskip
\item \textbf{SPUs} \cite{Appel2014}. Event-driven Process Chains (EPCs), which is the flowchart-based approach for BP modelling, is commonly used in ERP systems. In order to integrate IoT-driven ERP system in which the system needs to handle the IoT device generated event streams, Event Stream Processing Units (SPUs) project propose a middleware framework to translate the EPCs to the service-oriented BPMN-based system.
\\
The SPUs project provides a comprehensive solution in both model design and execution platform. In order to model the IoT-driven event streams, SPUs project introduces a number of new EPCs elements, together with the mapping approach between the EPCs elements and BPMN. Specifically, the new BPMN element---{\it event stream task} can represent the continuous event stream data processing task. The flow of such task proceeds by the boundary non-interrupt signals attached to the task. Ideally, such a design provides the flexibility for the system to handle the event stream independently. Further, for the model design, the project has also developed the extension of Software AG's ARIS Process Performance Manager platform for including the proposed BP model approach.
\smallskip
\item \textbf{Caraca{\c{s}} et al.}~\cite{Caracas2011,Caracas2012thesis}~
of IBM Zurich Research has proposed a highly integrated WSN-driven BPMS framework for HVAC in which the IoT/WSN devices are participating in BP based on executing the dynamically deployed tasks. In general, the project's theoretical BP modelling is focusing on WSN-driven workflow pattern design. The proposed patterns include: 
    \begin{enumerate}
    \item completion of asynchronous operation in WSN; 
    \item parallel starting asynchronous operation in WSN; 
    \item WSN task exception handling; 
    \item significant states of asynchronous interface; 
    \item addressing IoT/WSN devices in BPMN; 
    \item single and multiple messages receiving in WSN; 
    \item processes involve Time Division Multiple Access (TDMA) protocols. 
    \end{enumerate}
These patterns can serve as guidance for model designers who involve in WSN-driven workflow designs. Additionally, the authors have implemented the modelling design as the extension of the Web-based Oryx BPMN2.0 editor.
\end{itemize}

%
%
\begin{table*}[t]
\scriptsize
\centering
\caption{Comparison of IoT-driven Business Process Modelling Frameworks
}
\label{tb:modellingFrameworks}
\renewcommand{\arraystretch}{1.5}
\begin{tabular}{lll|ll|lllll|lllll}
\hline
\multicolumn{1}{l|}{\multirow{2}{*}{}} & \multicolumn{2}{l|}{Scope} & \multicolumn{2}{l|}{\pbox{20cm}{BP\\Operation}} & \multicolumn{5}{l|}{Modelling Methods} & \multicolumn{5}{l}{Modelling Tool} \\ \cline{2-15} 
\multicolumn{1}{l|}{}                  & IA          & IR          & CN              & DI             & EL    & MM    & PP    & SM    & SP    & SC    & OY    & BW   & AR   & PT   
\\ \hline
\multicolumn{1}{l|}{IoT-A}             &           &           &               &              & \checkmark     & \checkmark     &     &     &     & \checkmark     &     &    &    &    
\\ \hdashline[1pt/2pt]
\multicolumn{1}{l|}{makeSense}         
& \checkmark           
&           
& \checkmark               
& \checkmark              
& \checkmark     
&     
&     
&     
& \checkmark     
& \checkmark     
&     
&    
&    
&      
\\ \hdashline[1pt/2pt]
\multicolumn{1}{l|}{ASPIRE}            & \checkmark           &\checkmark           &  \checkmark             &              &     &     &     & \checkmark     & \checkmark     &     &     & \checkmark    &    &    
\\ \hdashline[1pt/2pt]
\multicolumn{1}{l|}{Sungur et al.}          &           &           &               &              & \checkmark     & \checkmark     &     & \checkmark     &     &     & \checkmark     &    &    &    
\\ \hdashline[1pt/2pt]
\multicolumn{1}{l|}{uBPMN}             & \checkmark           &           & \checkmark               &              & \checkmark     & \checkmark     &     & \checkmark     &     &     &     &    &    &    
\\ \hdashline[1pt/2pt]
\multicolumn{1}{l|}{SPUs}              & \checkmark           &           & \checkmark               &              & \checkmark     &     &     &     &     &     &     &    & \checkmark    &    \\ \hdashline[1pt/2pt]
\multicolumn{1}{l|}{Caraca{\c{s}} et al.}    & \checkmark           &           & \checkmark               & \checkmark              &     &     & \checkmark     &     &     &     & \checkmark     &    &    &    
\\ \hline
\multicolumn{15}{l}{\checkmark = subject addressed/supported; (blank) = not addressed/supported.}                                                                                               
\end{tabular}
\\
\medskip
\renewcommand{\arraystretch}{1.2}
\begin{tabular}{llllllll}
\hline
\multicolumn{2}{l}{Scope}    & \multicolumn{2}{l}{BP Operation}      & \multicolumn{4}{l}{BP Modelling Approach}                 \\ \hline
IA & Intra-organisational    & CN      & Centralised    & EL  & IoT element        & PP      & Process pattern      \\
IR & Inter-organisational    & DI      & Distributed    & MM  & Meta-model         & SP      & Specification        \\
   &                         &         &                             & SM  & Schema             &         &                      \\ \hline
\multicolumn{8}{l}{Modelling Tool}                                                                                               \\ \hline
SC & \multicolumn{3}{l}{Signavio GmbH Signavio Core Components}      & OY  & \multicolumn{3}{l}{Oryx BPMN editor}                \\
BW & \multicolumn{3}{l}{Business Process Workflow Management Editor} & PT  & \multicolumn{3}{l}{Proposing modelling editor tool} \\
AR & \multicolumn{3}{l}{Software AG - ARIS}                          &     & \multicolumn{3}{l}{}                                \\ \hline
\end{tabular}
\end{table*}
%
%

\smallskip
\noindent 
\textbf{Discussion}

\smallskip
Table~\ref{tb:modellingFrameworks} provides a comparison of the modelling frameworks. Specifically, modelling-based frameworks aim to introduce its elements in existing modelling standards such as BPMN. In general, a complete solution should include the graphical model approach, its associated meta-model and the model schema. Finally, the modelling approach needs to also provide a practical tool. For example, the approach can provide an extension of an existing BP model editor, which also can generate a machine-readable meta-model for further use in the {\it implement/configure phase}. As the table shows, most frameworks have provided the complete need of modelling frameworks. Moreover, the comparison also shows that most modelling frameworks only focus on the model for centralised BP execution and intra-organisational BPMS. Although two frameworks have addressed distributed BP execution, they have not fully addressed how to model the IoT-driven BP model in terms of representing the IoT tasks, data objects and events etc. Similarly, the only framework that involves inter-organisational BPMS also has not provided a solution for modelling the IoT elements in the inter-organisational level. These indicate the research gap of BPMS4IoT in modelling domain.

\subsection{System Integration}
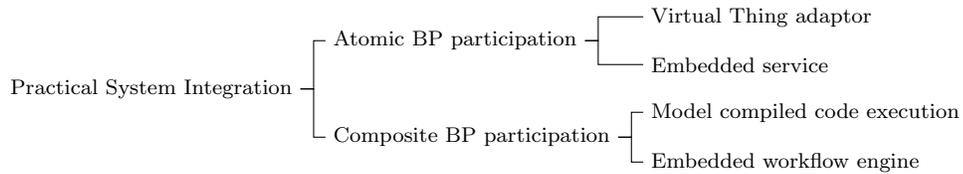
\begin{figure*}[h]
\footnotesize
\centering
\begin{forest}
  for tree={
    edge path={
      \noexpand\path[\forestoption{edge}](!u.parent anchor) -- +(5pt,0) |- (.child anchor)\forestoption{edge label};},
    grow=0,
    reversed, 
    parent anchor=east,
    child anchor=west, 
    anchor=west,
    if n children=0{tier=word}{}
}
[Practical System Integration,
	[\pbox{20cm}{Atomic BP participation},
		[Virtual Thing adaptor]
		[Embedded service]
	]
	[\pbox{20cm}{Composite BP participation},
		[Model compiled code execution]
		[Embedded workflow engine]
	]
]
\end{forest}
\\
\bigskip
\caption{Taxonomy of Integration Frameworks
}
\label{fig:taxonomyIntegrationFrameworks}
\end{figure*}

In general, we can classify the system integration frameworks into two taxonomies (see Figure~\ref{fig:taxonomyIntegrationFrameworks}): The \textit{atomic BP participation} denotes a model that allows process involving IoT devices as static operations such as HTTP/CoAP service invocations where the behaviours of the IoT devices cannot be re-programmed at runtime. On the other hand, in the \textit{composite BP participation} model, the behaviour of IoT devices can change dynamically at runtime based on the workflow model assigned to them. Furthermore, based on the discussion in Section~\ref{sec:BPE}, each of the taxonomy can further be split into two sub-taxonomies. We summarise the system integration frameworks in each sub-taxonomy as follows.

\medskip
\noindent
\textit{Virtual Thing Adaptor-based Integration}
\begin{itemize}
\item \textbf{Adventure} \cite{Schulte2014}. The goal of FP7 project---Adventure (ADaptive Virtual ENTerprise manufacturing Environment) is to develop a platform that can help manufacturing process as a virtual factory in the cloud. Considering the source of production will become highly distributed in the future, the platform utilises IoT technologies to assist the governance of the processes.
\\
With Adventure, the system creates each production plan as an instance managed in the cloud, in which the cloud instance of production proceeds with the BPMS that can discover and integrate the production system from client manufacturers.
\\
In order to deploy the BP model, the proposed Cloud Process Execution Engine (CPEE) will translate the BPMN to the proposed  Domain Specific Language (DSL), which specifies the code to express the workflow and is directly executable as Ruby language.
\\
Basically, the architecture of Adventure is following a general SOA model, in which the Cloud BPMS communicates with the IoT/WSN devices of different manufacturers via the flow of {\it Gateway-to-Adaptor}$\rightarrow${\it Adaptor to IoT/WSN devices}, which indicates that the expectation of IoT/WSN devices to provide or to install the adaptor software that can enable the communication. Particularly, an expectation of Adaptor can be the classic HTTP mobile Web servers.
\smallskip
\item \textbf{ASPIRE} \cite{Kefalakis2011}. The system integration of ASPIRE provides a runtime middleware---AspireRFID Programmable Engine (PE). Generally, the PE handles the low-level configuration between the central system and RFID applications. Further, it deploys the configuration generated from BPWME (BP editor) to the executable workflow for the system.
\smallskip
\item \textbf{SPUs} \cite{Appel2014}.~
As mentioned previously, SPUs propose the modelling approach as the extension in ARIS Process Performance Manager platform. Accordingly, the project has developed an ESB-based middleware---Eventlet, which can execute the output BPMN meta-data from ARIS. 
\\
Eventlet, which mainly communicates by WSDL/SOAP, provides the event stream filtering mechanism based on Java Message Service (JMS). Further, since the framework utilises ARIS, which is one of the well-known process mining tools~\cite{Aalst2015}, it indicates the potential extension for introducing system optimisation mechanisms in the future.
\smallskip
\item \textbf{GaaS} \cite{Wu2012}. Gateway as a Service (GaaS) project aims to propose a platform that can easily integrate heterogeneous IoT devices with Web-based BPMS. The architectural design of GaaS is based on cloud-centric SOA, which consists of three main layers:
   \begin{itemize}
    \item[$\circ$\;] {\it WoT Infrastructure layer}, which corresponds to IaaS that shares resources with third-party systems. In this layer, the front-end IoT devices connect with their own backend servers and the backend servers can join the WoT Infrastructure layer as Gateway services.
    \item[$\circ$\;] {\it Service and Business Operation layer} corresponds to PaaS, which handles the service composition and business process management. 
    \item[$\circ$\;] {\it Intelligent Service layer}, which corresponds to SaaS that provides application and UI to end users.
    \end{itemize}
The IoT device integration in GaaS derives from Valpas Gateway developed by Aalto University based on ThereGate. Additionally, ThereGate is a practical development of Nokia Home Control Center (HCC). It is a Linux-based platform with an open interface and a software engine called ThereCore, which integrates IoT devices operated on ZigBee/Z-Wave/Bluetooth protocols.
\end{itemize}

%
\medskip
\noindent 
\textit{Embedded Service-based Integration}
\begin{itemize}
\item \textbf{RWIS} \cite{Yang2012}. RESTFul geospatial Workflow Interoperation System (RWIS) project proposes a new platform that utilises Open Geospatial Consortium (OGC) Web Processing Services (WPS) for sensor devices to provide geographical composite services. RWIS utilises XPDL to describe BP models and the system will translate the XPDL to BPEL for execution.
\\
RWIS organises Sensor Planning Service (SPS) and Sensor Observation Service (SOS) as Sensor Information Accessing (SIA) workflow. Further, it deploys OGC’s Web Processing Service (WPS) and OGC’s Web Coverage Service (WCS) Sensor Information Processing (SIP) workflow. Generally, the system achieves the interoperation in OpenWFE-based SIA workflow and BPEL-based SIP workflow. From the practical perspective, RWIS is focusing on integrating IoT with BPMS fully based on the compliance of OGC’s Web service standards.
\smallskip
\item \textbf{Decoflow} \cite{Loke2003} is a service-oriented WfMS proposed for home automation. Decoflow project introduces a modelling language called DySCo for modelling the abstract BP workflow. Fundamentally, the Decoflow runtime system is based on service-oriented BPEL4WS, in which the communication between the management system and devices relies on the embedded SOAP Web services hosted on the devices.
\\
The project also proposes a graphical design tool, which can generate and validate  BPEL meta-model for execution. Further, the extension of the project introduced a {\it fault tolerant scheme} in which the system solves the runtime device failure by using context-aware failure substitution. The faulty device will be replaced  by substitution \cite{Chang2008} based on the same class of context they can provide.
\end{itemize}

\medskip
\noindent 
\textit{Model Compiled Code Execution-based Integration}
\begin{itemize}
\item \textbf{makeSense} \cite{Casati2012,Tranquillini2012,Tranquillini2015} provides {\it model-to-execution} mechanisms for the system integration. It mainly supports three approaches: 
\begin{enumerate}
\item The proposed BPMN4WSN compiler will generate the IoT/WSN device executable binary code from the meta-model generated from the proposed BPMS4WSN editor. 
\item If the IoT/WSN device is powerful enough to embed Process Engine (i.e. for executing workflow model), the BPMN4WSN compiler will generate an executable workflow for the device; 
\item If the IoT/WSN device is communicated via the proxy device, the BPMN4WSN compiler will generate proxy configuration and send the configuration meta-data to the proxy device. 
\end{enumerate}
The makeSense project has tested the prototype on Contiki devices that can execute the output from the proposed BPMN4WSN compiler. Overall, makeSense is suitable for the systems that require distributed and dynamic deployed BP workflow execution at the edge network of IoT systems.
\smallskip
\item \textbf{Caraca{\c{s}} et al.}~\cite{Caracas2011,Caracas2012thesis} proposes a model compiler middleware that can translate the editor-generated BP model to executable program source code in C\# or Java language, which is executable by IBM Mote Runner OS-based devices.
\\
Further, consider that the runtime deployment-based BP participation can influence the power consumption of IoT/WSN devices, the project also proposes a strategy to reduce the power consumption based on the optimised sleeping scheme. Generally, the scheme utilises the machine learning algorithm that learns from the power consumption of the wake-up on IoT/WSN devices and autonomously reconfigures the process.
\smallskip
\item \textbf{Presto} \cite{Giner2010} is a pluggable software architecture for leveraging Tag-based IoT technologies with BPMS. It focuses on supporting human workers' participation in the BP in which the mobile devices serve as the medium between human works and the system. Fundamentally, Presto's primary role is to assist the human worker-involved workflow task allocation. It can automatically allocate tasks to workers depending on certain contextual factors such as how many pending tasks exist at the worker's to-do-list. 
\\
Presto also proposed a middleware platform---Parkour, which is capable of translating the Parkour model (a customised modelling specification) to Ecore meta-model (part of the Eclipse Modelling Framework). Afterwards, the Ecore meta-model can be translated to the execution languages of a specific platform such as Java, Android or iOS.
\smallskip
\item \textbf{LTP} \cite{Glombitza2011}. The Lean Transport Protocol (LTP) project introduces a middleware framework to integrate IoT devices with BPMS with the enhancement of the proposed new protocol stack. In this project, IoT devices communicate with BPEL system using LTP-enhanced messages. Additionally, LTP provides SOAP message compression (SMC) to improve the communication performance in the edge network of the IoT system.
\\
Overall, the project facilitates the BP participation from IoT devices by utilising the dynamic model compiled code deployment approach in which the BPEL workflow model is compiled into C++ code. Afterwards,  the IoT devices that have embedded GNU Compiler Collection (GCC) C++ compilers can execute the compiled code.
\end{itemize}

\medskip
\noindent 
\textit{Embedded Workflow Engine-based Integration}
\begin{itemize}
\item \textbf{MOPAL} \cite{Peng2014}. MCC system is a common approach used in healthcare application where physicians can remotely allocate a sequential task to the healthcare nurses via their mobile devices, which act as mediums. Commonly, such application is mainly relying on the cloud-side server for deploying, executing and managing the entire BP workflow. However, it also faces the problem derived from the unstable mobile Internet connection.
\\
MOPAL project aims to solve the problem by introducing the disconnected workflow execution approach in which the physician-defined workflow can be dynamically executed on the nurses' mobile device without the need of the Internet. In general, the architectural design of MOPAL is following the common SOA that utilises the native components of mobile devices (e.g. camera, microphone) as services. Hence, the embedded workflow engine on the mobile device can access the native components seamlessly. 
\\
The framework also contains a customised BPMN workflow developed based on XPath and SQLite. The project has implemented the MOPAL prototype  for Android OS. In general, MOPAL decouples the BP execution between the distant cloud servers and the front-end mobile mediums.
\\
Further, for the runtime management, MOPAL defines two context constraints to reduce the runtime failure caused by the context influences. The first class of the constraints is the assignment constraints, which is based on the matchmaking between the workflow task assignment and the profile and context information of the candidate participant. The second class of the constraints is the execution constraints, which are related to the context of the physical world such as current geographic location and time.
\smallskip
\item \textbf{Dar et al.} \cite{Dar2015} AAL applications often utilise smartphones as mediums to interact with the heterogeneous front-end RESTful service-based IoT environments. In order to fully support the need for highly dynamic workflow deployment, the system requires the mechanism for dynamically deploying and executing the workflows on smartphones. 
\\
Dar et al. propose a framework for enabling RESTful service integration-based workflow system with the feature of the  choreography support. In this case, the system can distribute the workflow to different front-end IoT devices for execution. 
\\
The project has implemented a prototype on Android OS device based on the ported version of Activiti BPM engine. Additionally, the framework includes common protocols used in IoT systems such as CoAP for constrained service invocation and MQTT for event stream subscription.
\\
Furthermore, the event stream mechanism enhances the runtime device failure detection. The system monitors the health of the BP execution by utilising the MQTT-based periodical reporting method. If the system detects the failure, it is capable of performing substitution by re-assigning the workflow to a different device for execution.
\\
Overall, the choreography feature provided by this project reduces the need for the frequent transmission between front-end BP activities and the distant cloud, which is also a well-studied approach in MCC field \cite{Chang2015scc}.
\smallskip
\item \textbf{SCORPII} \cite{Chang2015scc}. In IoT technology assisted smart urban area, heterogeneous devices are discoverable and they can provide various information. When an Ambient Assisted Living (AAL) application relies on the information provided by the surrounding IoT devices, it faces the challenge of how to rapidly discover the devices that can provide relevant information in a timely manner?
\\
In order to resolve the question, SCORPII project provides a middleware framework based on dual BPEL workflow execution engines hosted on the dynamically launched utility cloud and the user's mobile device. The workflow system is capable of dynamically assigning the service discovery tasks between cloud and mobile in order to achieve the best cost-performance efficiency.
\\
In detail, SCORPII’s mobile-embedded BPEL workflow execution engine is a customised engine developed for iOS devices. The engine utilises  GDataXML (XPath) and CocoaHTTPServer for the practical implementation. The prototype supports the main operations of BPEL such as sequential and parallel task executions.
\\
Further, for the runtime management, SCORPII supports the resource-aware cost-performance index scheme to identify the most efficient way of executing the workflow tasks between the cloud and the mobile device. Generally, the decision-making is based on the environmental factors such as how large the service description data from the distant discovery servers is involved.
\end{itemize}


\noindent 
\textit{Other Related Frameworks}

\smallskip
\noindent
Beside the frameworks described above, there are other relevant frameworks consist with little involvement of BPMS. Following is the summary of these frameworks. Note that the comparison table does not include these frameworks.
\begin{itemize}
\item \textbf{EBBITS} \cite{Furdik2013} is an FP7 project that mainly focuses on CPS integration using LinkSmart SDK (\url{https://linksmart.eu/}). Accordingly, the project plans to utilise BPMS for orchestration with the RESTful service-embedded IoT devices. EBBITS refers the work of IoT-A for their BPMS-related components.
\smallskip
\item \textbf{VITAL} \cite{VITAL} is an FP7 smart city project that proposes a new domain-specific language-based BP modelling and configuration approach. The approach aims to introduce the modelling language as a new specification for the citywide system integration that can support the composition among inter-organisational IoT management systems.
\smallskip
\item \textbf{edUFlow} \cite{Jung2012} is a practical integration framework that composes the open-source Esper event correlation engine and GlassFish Message Queue to enable an IoT-driven event stream processing platform. The project also provides a GUI-based editor for the process configuration and runtime monitoring.
\end{itemize}

Table~\ref{tb:comparisonIntegration} illustrates an overview of the existing BPMS4IoT integration frameworks. Based on the characteristics from Mobile Cloud Computing perspective, these frameworks may fulfil different needs in integrating mobile/wireless IoT devices into BPMS.

\begin{sidewaystable*}
\caption{Comparison of System Integration Frameworks}

\label{tb:comparisonIntegration}
\footnotesize
\renewcommand{\arraystretch}{1.2}
\begin{tabular}{|l||l|l|l|l|l|l|l|}
\hline 
                        & \textbf{\begin{tabular}[c]{@{}l@{}}Domain /\\ Use case\end{tabular}} & \textbf{\begin{tabular}[c]{@{}l@{}}Integration\\ Model\end{tabular}} & \textbf{\begin{tabular}[c]{@{}l@{}}IoT Device\\ Integration\\ Platform /\\ Protocol\end{tabular}} & \textbf{\begin{tabular}[c]{@{}l@{}}BP Model\\ Standard\end{tabular}} & \textbf{\begin{tabular}[c]{@{}l@{}}BP Execution\\ Engine\end{tabular}}           & \textbf{\begin{tabular}[c]{@{}l@{}}Prototype\\ IoT Device /\\ Platform\end{tabular}} & \textbf{\begin{tabular}[c]{@{}l@{}}Runtime\\ Management\end{tabular}}   
\\ \hline \hline
\textbf{Adventure}      & Manufacturing                                                        & \begin{tabular}[c]{@{}l@{}}Virtual Thing\\ adaptor\end{tabular}      & Cloud-centric WS                                                                                & BPMN                                                                 & \begin{tabular}[c]{@{}l@{}}Customised\\ BPMN engine\end{tabular}                 & ---                                                                                  & ---                                                                     
\\ \hdashline[1pt/2pt]
\textbf{ASPIRE}         & Logistics                                                            & \begin{tabular}[c]{@{}l@{}}Virtual Thing\\ adaptor\end{tabular}      & ASPIRE PE                                                                                       & XPDL                                                                 & \begin{tabular}[c]{@{}l@{}}BPWME \& Aspire-\\RFID middleware\end{tabular}         & RFID                                                                                 & ---                                                                     
\\ \hdashline[1pt/2pt]
\textbf{SPUs}           & Logistics                                                            & \begin{tabular}[c]{@{}l@{}}Virtual Thing\\ adaptor\end{tabular}      & SOAP                                                                                            & BPMN                                                                 & \begin{tabular}[c]{@{}l@{}}Software AG\\ ARIS platform\end{tabular}              & EPC/RFID                                                                             & \begin{tabular}[c]{@{}l@{}}Event stream\\ filtering\end{tabular}        
\\ \hdashline[1pt/2pt]
\textbf{GaaS}           & \begin{tabular}[c]{@{}l@{}}AAL/\\ Healthcare\end{tabular}            & \begin{tabular}[c]{@{}l@{}}Virtual Thing\\ adaptor\end{tabular}      & ThereGate                                                                                       & BPEL                                                                 & Apache ODE                                                                       & \begin{tabular}[c]{@{}l@{}}ZigBee/Z-wave\\ sensors\end{tabular}                      & ---                                                                     
\\ \hdashline[1pt/2pt]
\textbf{RWIS}           & \begin{tabular}[c]{@{}l@{}}Location-based\\ service\end{tabular}     & \begin{tabular}[c]{@{}l@{}}Embedded\\ service\end{tabular}           & OGC WS                                                                                          & \begin{tabular}[c]{@{}l@{}}XPDL \&\\ BPEL\end{tabular}               & RWIS platform                                                                    & OGC sensors                                                                          & ---                                                                     
\\ \hdashline[1pt/2pt]
\textbf{Decoflow}       & Smart home                                                           & \begin{tabular}[c]{@{}l@{}}Embedded\\ service\end{tabular}           & SOAP WS                                                                                         & BPEL                                                                 & GlassFish                                                                        & ---                                                                                  & ---                                                                     
\\ \hdashline[1pt/2pt]
\textbf{makeSense}      & HVAC                                                                 & \begin{tabular}[c]{@{}l@{}}Compiled\\ code$^{\dagger}$\end{tabular}              & Contiki C                                                                                       & BPMN                                                                 & \begin{tabular}[c]{@{}l@{}}Model compiled\\ C code execution\end{tabular}        & \begin{tabular}[c]{@{}l@{}}Contiki OS\\ devices\end{tabular}                         & ---                                                                     \\ \hdashline[1pt/2pt]
\textbf{Caracas et al.} & HVAC                                                                 & \begin{tabular}[c]{@{}l@{}}Compiled\\ code$^{\dagger}$\end{tabular}              & IBM Mote Runner                                                                                 & BPMN                                                                 & \begin{tabular}[c]{@{}l@{}}Model compiled C\#/\\Java code execution\end{tabular} & \begin{tabular}[c]{@{}l@{}}Mote Runner\\ VM/OS devices\end{tabular}                  & \begin{tabular}[c]{@{}l@{}}Energy\\ efficiency\end{tabular}             
\\ \hdashline[1pt/2pt]
\textbf{Presto}         & Smart business                                                       & \begin{tabular}[c]{@{}l@{}}Compiled\\ code$^{\dagger}$\end{tabular}              & OSGi                                                                                            & ---                                                                  & Parkour / Java OSGi                                                              
& \pbox{20cm}{OSGi on\\Android}                                                                      & ---                                                                     
\\ \hdashline[1pt/2pt]
\textbf{LTP}            & Logistics                                                            & \begin{tabular}[c]{@{}l@{}}Compiled\\ code$^{\dagger}$\end{tabular}              & LTP \& SOAP                                                                                     & BPEL                                                                 & \begin{tabular}[c]{@{}l@{}}Model compiled\\ C++ code execution\end{tabular}      
& \pbox{20cm}{GCC on\\Android}                                                                       & ---                                                                     
\\ \hdashline[1pt/2pt]
\textbf{MOPAL}          & \begin{tabular}[c]{@{}l@{}}AAL/\\ Healthcare\end{tabular}            & \begin{tabular}[c]{@{}l@{}}Embedded\\ WF engine$^{\ddagger}$\end{tabular}         & SOAP                                                                                            & BPMN                                                                 & \begin{tabular}[c]{@{}l@{}}Customised BPMN\\ engine\end{tabular}                 & Android                                                                              & \begin{tabular}[c]{@{}l@{}}Constraint\\ execution\end{tabular}          
\\ \hdashline[1pt/2pt]
\textbf{SCORPII}        & \begin{tabular}[c]{@{}l@{}}AAL/\\ Healthcare\end{tabular}            & \begin{tabular}[c]{@{}l@{}}Embedded\\ WF engine$^{\ddagger}$\end{tabular}         & RESTful HTTP                                                                                    & BPEL                                                                 & \begin{tabular}[c]{@{}l@{}}Customised BPEL\\ engine\end{tabular}                 & Android                                                                                  & \begin{tabular}[c]{@{}l@{}}Cost-performance\\ balancing\end{tabular}    
\\ \hdashline[1pt/2pt]
\textbf{Dar et al.}     & \begin{tabular}[c]{@{}l@{}}AAL/\\ Healthcare\end{tabular}            & \begin{tabular}[c]{@{}l@{}}Embedded\\ WF engine$^{\ddagger}$\end{tabular}         & \begin{tabular}[c]{@{}l@{}}RESTful HTTP/\\ CoAP/MQTT\end{tabular}                               & BPMN                                                                 & \begin{tabular}[c]{@{}l@{}}Ported Activiti BPM\\ engine\end{tabular}             & Android                                                                              & \begin{tabular}[c]{@{}l@{}}Reactive failure\\ substitution\end{tabular} \\ \hline
\end{tabular}
$^{\dagger}$ Model compiled code execution\\
$^{\ddagger}$ Embedded workflow engine
\end{sidewaystable*}

\medskip
\noindent 
\textbf{Discussion}

\smallskip
\noindent
The comparison indicates that the Virtual Thing Adaptor-based integration model is commonly applied in the large scope enterprise systems such as manufacturing and logistics where the primary purpose of the integration is to trace the items. In general, the IoT-driven manufacturing and logistics systems only utilise the fairly simple IoT device such as RFID/RFID readers or simple function sensors.

Embedded service-based approaches deploy the IP-based Web services on devices to enable the common service invocation between the {\it smart devices} and the management system. Generally, this model is fully compliant to BPEL4WS-based BPMS and it usually applies to the location-based systems or the smart home systems.

In the case of a large-scale IoT-driven system such as HVAC or smart building system where the system requires a certain level of self-management and self-configuration, there is a need to provide dynamic process configuration on the resource constrained IoT devices. However, IoT-devices used in HVAC or smart building systems are usually having constrained resources, in which implementing the standard-compliant workflow execution engines on them is not performance efficient. Therefore, the model compiled code execution-based approach becomes the promising solution. Accordingly, the model compiled code execution-based frameworks usually provide a complete solution that covers from BP modelling tool to the  deployment of the runtime system.

As Table~\ref{tb:comparisonIntegration} shows, all the projects that utilised the embedded workflow engine-based approaches are originally developed for AAL scenarios. Explicitly, such scenarios utilise smartphones as mediums for interacting with the IoT-driven ubiquitous environments and they require the on-demand timely response and also the choreography-based service composition. Hence, the embedded workflow engines become a feasible option.

According to the study of this section, current frameworks are in the early stage since they have not explicitly addressed optimisation in a BPMS. Optimisation involves scalability and continuously improve the BP model using techniques like process mining or process discovery etc.

Further, only a few frameworks have considered fault tolerance and context-awareness in the runtime management of BPMS4IoT. At this stage, it is not clear whether the existing process models can already address the fault tolerance and context-awareness or not, especially when the IoT system involves mobile objects. Since the BPMS4IoT projects commonly design their BPMS for the static participant and controlled environment, integrating BPMS with IoT can raise new issues. For instance, addressing fault tolerance requires the process model to describe the detail process of the involved entities. The BP model that does not address the activities of the IoT entities (e.g. sensor, actuator, reader etc.), may not be able to identify the cause of the failure, and hence, the BP modeller cannot design the corresponding recovery processes for the potential failures.

\subsection{Application Comparability Comparison}
\begin{figure*}[htbp]
    \centering
    \begin{subfigure}[t]{0.27\textwidth}
        \includegraphics[width=\textwidth]{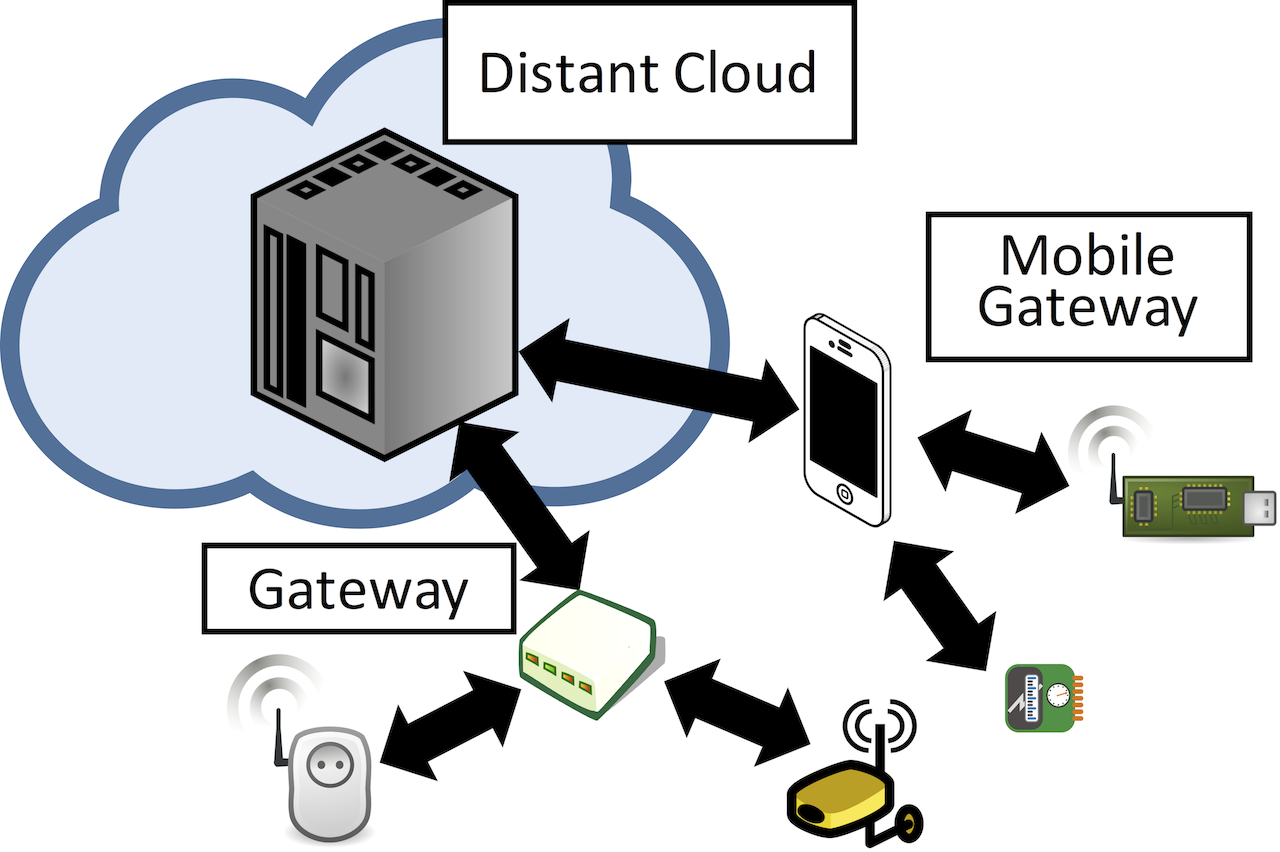}
        \caption{Distant data centre}
        \label{fig:model_cloud}
    \end{subfigure}
    \hfill%
    \begin{subfigure}[t]{0.35\textwidth}
        \includegraphics[width=\textwidth]{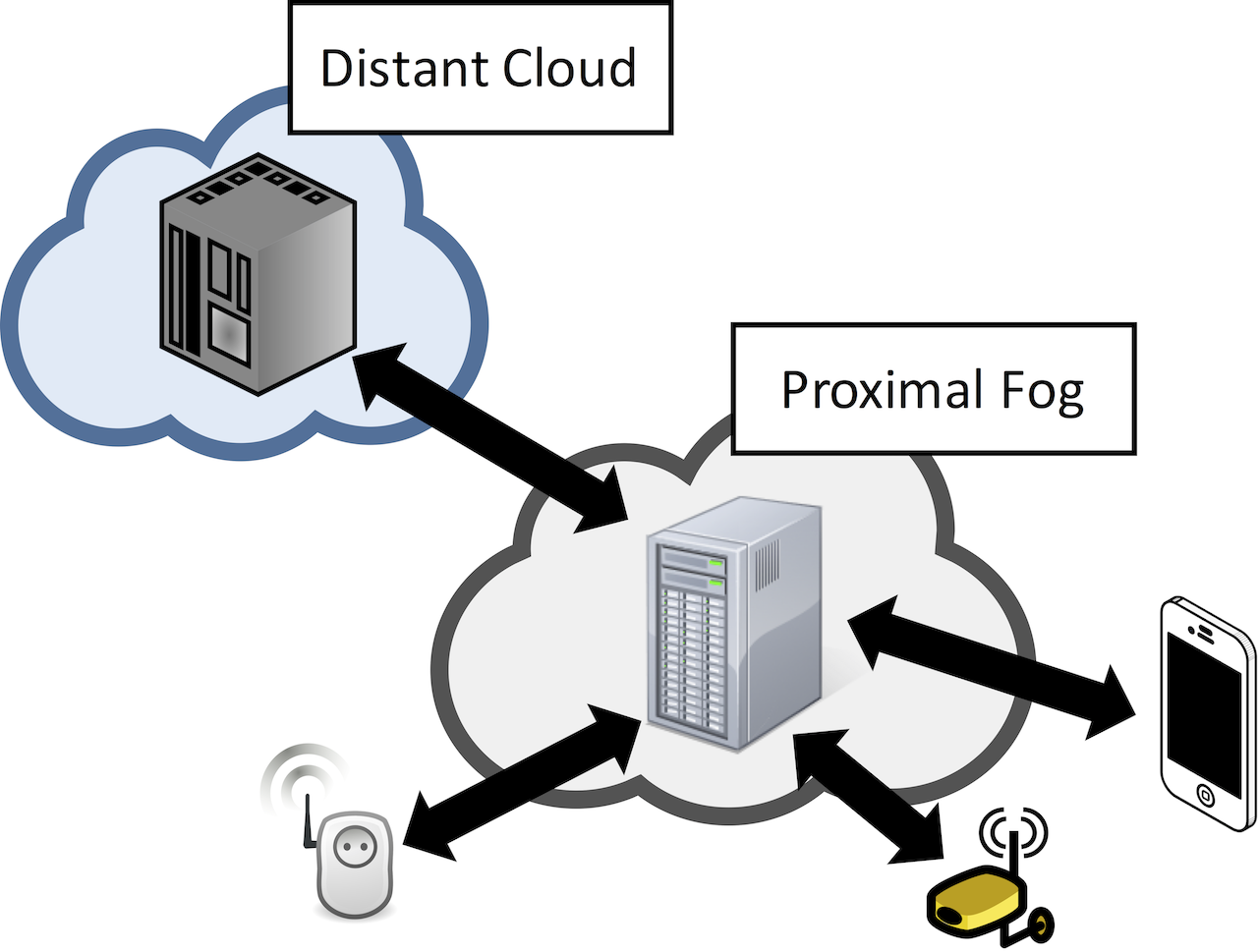}
        \caption{Fog computing}
        \label{fig:model_fog}
    \end{subfigure}
    \hfill%
    \begin{subfigure}[t]{0.35\textwidth}
        \includegraphics[width=\textwidth]{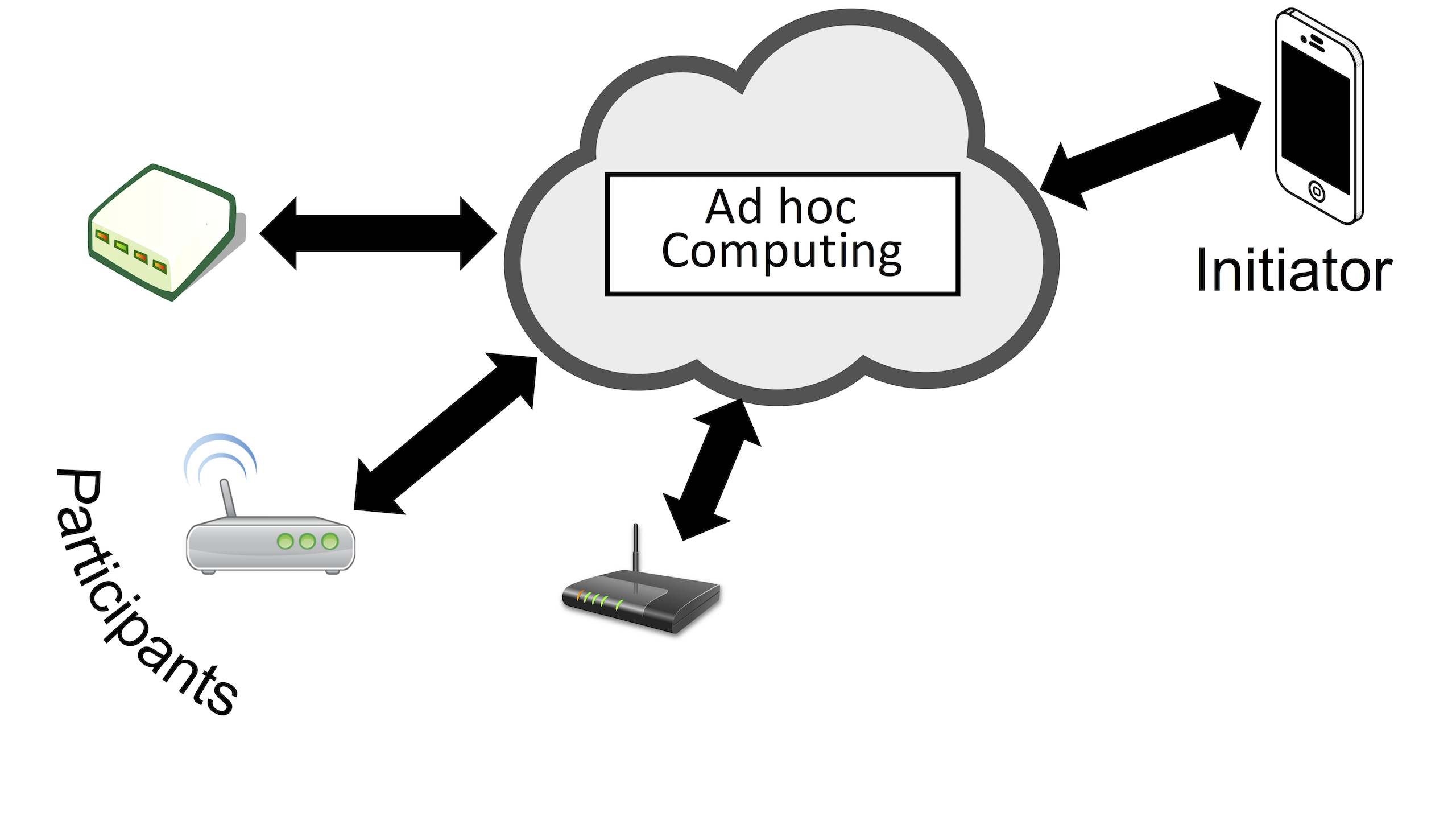}
        \caption{Ad hoc computing}
        \label{fig:model_mist}
    \end{subfigure}
    \\
    \begin{subfigure}[t]{0.27\textwidth}
        \includegraphics[width=\textwidth]{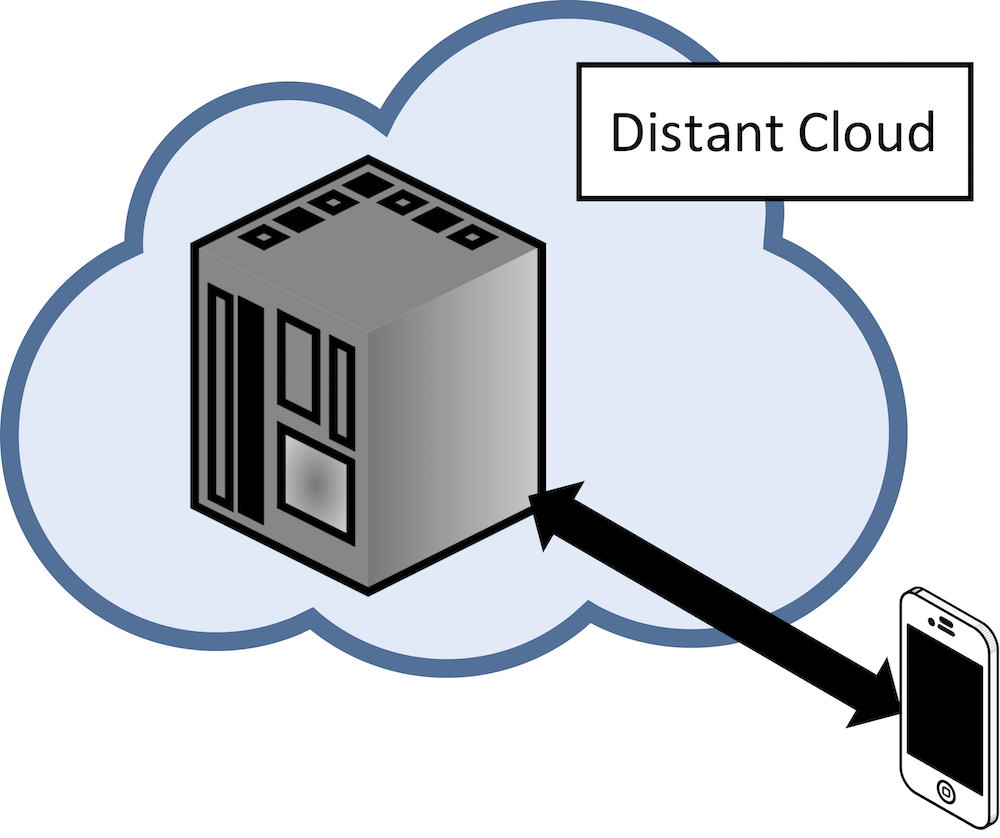}
        \caption{Distant Mobile Cloud}
        \label{fig:mcc_distant}
    \end{subfigure}
    \hfill%
    \begin{subfigure}[t]{0.32\textwidth}
        \includegraphics[width=\textwidth]{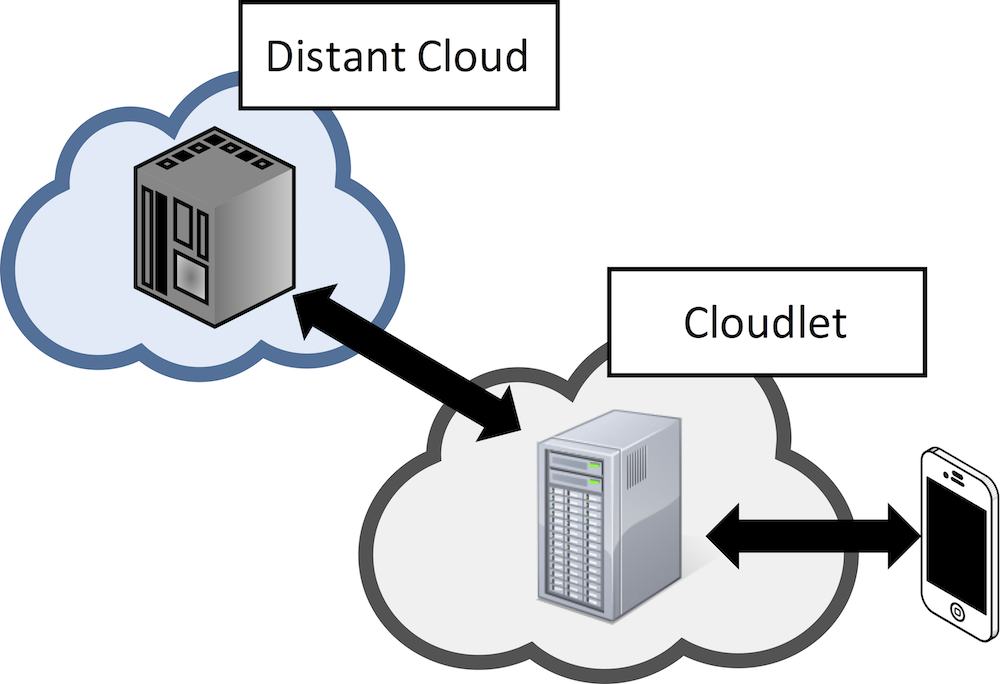}
        \caption{Mobile Edge Cloud}
        \label{fig:mcc_fog}
    \end{subfigure}
    \hfill%
    \begin{subfigure}[t]{0.37\textwidth}
        \includegraphics[width=\textwidth]{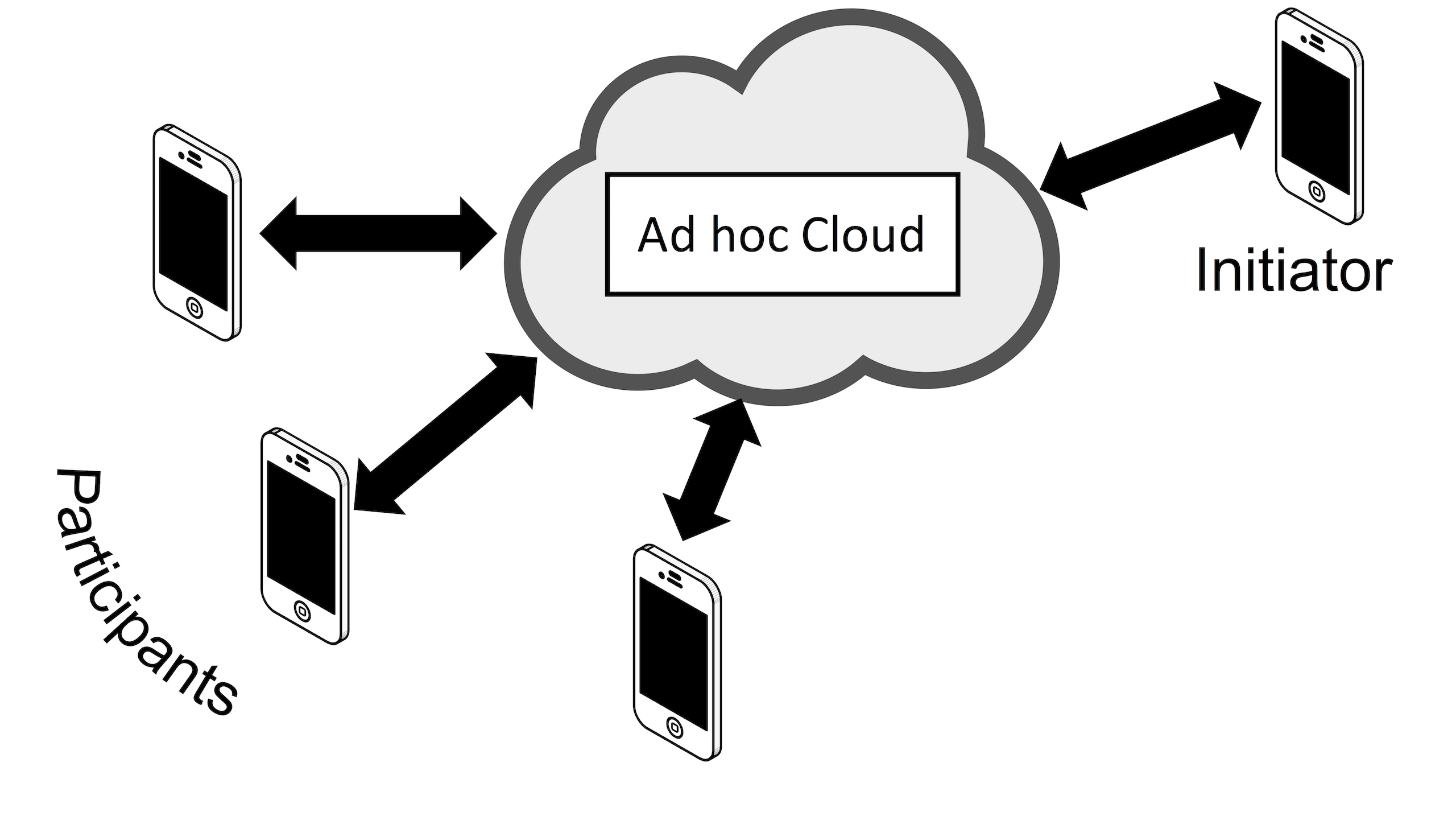}
        \caption{Mobile Ad hoc Cloud}
        \label{fig:mcc_adhoc}
    \end{subfigure}
    \caption{Correlative MCC and IoT deployment models. Note: the mobile devices can be replaced by any type of IoT devices.}
    \label{fig:mcc}
\end{figure*}

In this section, we try to identify the feasibility of the existing BPMS4IoT framework when we apply them to recent IoT systems. Generally, we can classify the recent IoT systems architecture into three main deployment models, which are fundamentally similar to the three MCC model types which has been discussed in \cite{Fernando2013}. The three models are (1) Distant Data Centre (Fig.~\ref{fig:model_cloud}), which refers to the Distant Mobile Cloud model (Fig.~\ref{fig:mcc_distant}); (2) Fog computing \cite{Bonomi2012} (Fig.~\ref{fig:model_fog}), which refers to the Mobile Edge Cloud model (Fig.~\ref{fig:mcc_fog}); (3) Ad hoc Computing \cite{Kortuem2010} (Fig.~\ref{fig:model_mist}), which  refers to Mobile Crowd Computing \cite{Loke2015} (Fig.~\ref{fig:mcc_adhoc}).

\bigskip
%
%

We summarise the three IoT models as follow.
\begin{itemize}
\item {\bf Distant data centre} is a classic model in which the cloud-side representations (e.g. virtual thing adaptor) of the IoT/Mobile  devices are delegating the BP for the devices. Most BPMS4IoT frameworks apply this model.
\smallskip
\item {\bf Fog computing} refers to mobile edge computing \cite{Patel2014} where the system utilises the MCC-driven {\it cloudlet} \cite{Satyanarayanan2009} concept. Generally, a {\it cloudlet} is a Virtual Machine (VM)-enabled server machine co-located in a cellular base station or a WLAN access point (e.g. CISCO Grid router). Notably, it is one of the major models of IoT system, which is to fulfil  the need of the rapid response from location-based ubiquitous sensing and context recognition processes. It is foreseeable that the future BPMS4IoT will highly involve in this model. For instance, the {\it cloudlet} can compile the BP model, encode/decode the message, handle the runtime monitoring and process event stream. Further, it can work as the middleware or adaptor between the distant cloud and the front-end mobile nodes. 
\smallskip
\item {\bf Ad hoc computing} can refer to Mobile Crowd Computing \cite{Loke2015} and Mist computing \cite{Pulli2011,Preden2015,Martin2015}, which focuses on utilising the edge network devices as the process participants. Generally, this model enables the devices to communicate with each other and collaboratively perform the tasks together. Such cooperated business process distribution requires a highly flexible execution mechanism. Hence, embedded workflow engine is the basic requirement to meet the high automation need.  
\end{itemize}

\begin{table}[h]
\centering
\caption{Compatibility in MC-BPMS4IoT models.}
\label{tb:compatibility}
\footnotesize
\begin{tabular}{|l|c|c|c|}
\hline
               & \multicolumn{1}{l|}{Distant Data Centre} & \multicolumn{1}{l|}{Fog Computing} & \multicolumn{1}{l|}{Ad hoc Computing} \\ \hline
IoT-A          & \checkmark                                        &                                        &                                          \\ \hline
Sungur et al.  & \checkmark                                        &                                        &                                          \\ \hline
uBPMN          & \checkmark                                        &                                        &                                          \\ \hline
Adventure      & \checkmark                                        &                                        &                                          \\ \hline
ASPIRE         & \checkmark                                        &                                        &                                          \\ \hline
SPUs           & \checkmark                                        &                                        &                                          \\ \hline
GaaS           & \checkmark                                        &                                        &                                          \\ \hline
RWIS           & \checkmark                                        &                                        &                                          \\ \hline
Decoflow       & \checkmark                                        &                                        &                                          \\ \hline
makeSense      & \checkmark                                        & \checkmark                                     &                                          \\ \hline
Caracas et al. & \checkmark                                        & \checkmark                                     &                                          \\ \hline
Presto         & \checkmark                                        & \checkmark                                     &                                          \\ \hline
LTP            & \checkmark                                        & \checkmark                                     &                                        \\ \hline
MOPAL          & \checkmark                                        & \checkmark                                     & \checkmark                                       \\ \hline
SCORPII        & \checkmark                                        & \checkmark                                     & \checkmark                                       \\ \hline
Dar et al.     & \checkmark                                        & \checkmark                                     & \checkmark                                       \\ \hline
\end{tabular}
\end{table}

Table~\ref{tb:compatibility} illustrates the compatibility between the existing BPMS4IoT frameworks and IoT system models. This classification is based on the design of their system architectures and deployment approaches. Since the {\it ad-hoc computing} model requires flexible and timely process execution, the frameworks that provide embedded workflow engines are more suitable than the others. The {\it Fog Computing} model requires distributed process execution to Fog nodes situated in the close proximity of the requesters. Therefore, the frameworks that provide the middleware for deploying model-compiled code to IoT device, or the frameworks that utilising embedded workflow engines are more feasible. As the table shows, most frameworks are compatible with the classic {\it distant data centre} model because they either lack the capability for distributing processes or their architectural designs and BP model designs have not considered the applications based on Fog or mobile ad-hoc environments.

\section{Potential Issues and Open Challenges}
Based on our study, in this section, we identify a number of challenges that have not yet been fully addressed in existing BPMS4IoT frameworks. 
\subsection{Challenges in (Re)design}
%
\subsubsection{IoT-driven Business Process Model Standardisation}
%
%
Currently, there is no common agreed way to model the IoT entities and their activities in business process model tools. Beside the different devices that require different ways of modelling (e.g. modelling actuator is different to modelling data collector), different projects have their own perspective in modelling the same type entities. For example, based on the review of this paper, there are at least four different approaches to model the sensor devices. Further, there is a need to model the IoT devices in detail in order to ease the procedure of transforming the process model into the machine-readable metadata form for the execution. Further, the need of the standard model in BPMS4IoT is not only about the notations, it also involves the meta-model level. Ideally, it is better to have the capabilities for the BP modeller to define what kind of protocol and operation the IoT devices use and how the processes perform the message flow. Such a need is required for BPMS that need to compose information from different sources dynamically.

The potential approach in this domain may compose the modelling standards \cite{Ko2009} with the recent design described in Section~\ref{sec:modellingThings} to propose a generic specification that is applicable in different IoT use cases.

\subsubsection{Hybrid Computational Process Architecture}
%
%
The performance-related challenge in IoT motivated various edge network computing models. Specifically, the Fog Computing model \cite{Bonomi2012}, which utilises the VM-enabled grid router machine to replace the partial mechanism of the distant cloud services, and Mobile Edge Computing model \cite{Patel2014}, which utilises the server machine co-located with the cellular network base stations as the computational resources. Furthermore, the Mist Computing model \cite{Pulli2011,Preden2015,Martin2015} or the Mobile Crowd Computing model \cite{Loke2015}, which utilises proximal IoT/Mobile devices as the resources for computational offloading, is also showing as the promising approach in improving the performance of IoT systems. Since all these models involve vast devices that operated with their own OS independently, composing these models in BPMS4IoT requires a more technical design to support the reliability of the system because of the dynamic nature of the heterogeneous mobile network environment.

\subsubsection{BPM in the Large}
%
%
BPMS4IoT shares similar {\it design phase} challenges with {\it BPM-in-the-Large} \cite{Houy2010}. For instance, the complexity derived from the large scope of inter-organisational BPMS4IoT requires the extensive process models that involve situation-awareness. In other words, the BP models need to be self-adaptive in different situations based on the stakeholders. Specifically, the challenge in the large scope system involves an adaptive BP model design and runtime management solution for the inter-organisational BPMS4IoT. Accordingly, existing BP model frameworks have not addressed this domain.

%
\subsection{Challenges in Implement/Configure}
\subsubsection{Heterogeneity}
%
%
The heterogeneity of the participative IoT devices in BPM involves the challenges in connectivity, discoverability, mobility/accessibility, self-management and self-configuring. Here, we discuss the heterogeneity of participants in different layers.
\begin{itemize}
\item \textit{Devices}. Although different IoT devices have different capabilities, but they may achieve the same purposes. Hence, BPMS4IoT requires an efficient ontology model to classify the IoT devices. For example, both of the modern high-end smartphone and the low power Raspberry Pi can perform the sensory data collection tasks. On the other hand, since their computational power is quite different, it can affect the overall performance of the process. Therefore, the BPMS need to clearly define the IoT entities.

\smallskip
\item \textit{Embedded Service}. Services provided by the IoT devices depend on the communication protocol such as classic Bluetooth, Bluetooth Smart~
LE, CoAP, Alljoyn (\url{https://allseenalliance.org}), MQTT, AMQP (\url{https://www.amqp.org}) etc. These common IoT protocols work in different ways, and they influence the performance of the management system. Future BPMS for IoT requires an adaptive solution to provide self-configuration among the different connectivity.

\smallskip

\item \textit{Service Description}. Existing frameworks have introduced various approaches to interacting with the IoT entities. For examples, Web service-oriented frameworks have applied WSDL, WADL \cite{Hadley2006}, DPWS \cite{Driscoll2009DPWS}; frameworks that focus on WSN have applied 
SensorML (\url{http://www.ogcnetwork.net/SensorML}) or SenML \cite{Jennings2012}; frameworks that concern about energy efficiency have utilised IETF CoAP (RFC7252) \& CoRE standards. Further, W3C has recommended JavaScript Object Notation for Linked Data standard for describing the IoT entities. Consequently, it is now becoming impossible to rely on one single standard to enable autonomous Machine-to-Machine (M2M) communication in IoT due to the lack of a global common standard for machine-readable metadata.

\smallskip
\item \textit{Service Discovery}. In IoT, relying on a global central repository is less possible. It is foreseeable that in the future, IoT systems will rely on a large scale federated service discovery network established on a mesh network topology. Moreover, the proximity-based and opportunistic discovery also stands for an important role to map the physical visibility with the digital visibility. The use cases of the Inter-organisational BPM, crowdsourcing, crowdsensing, social IoT, real-time augmented reality and the ambient assistance for mobile healthcare will all need such a feature to support the rapid establishment of M2M connection and collaboration in physical proximity, which indicates that the BP model design needs to consider the adaptive service discovery.

\end{itemize}

\subsubsection{Urban Computing}
%
%

The future IoT in a public area will also involve collaborative sensing network \cite{Salim2015,Loke2015,Loke2015ubicomp} in which different organisations or individuals may share the already deployed resources such as sensors and sensory data collector devices \cite{Wu2012}. Therefore, the process models for such an environment will become very complex and will face many challenges such as privacy and trust issues. The privacy issue involves what resources can be shared and how they will be shared? The trust issue involves how does an organisation provide the trustworthy resource or sensory data sharing? These issues will further require adaptive Quality of Service (QoS) models and the scalable Service Level Agreement (SLA) schemes to adapt to different situations and preferences. In summary, the collaborative IoT environments need to address the following challenges.

\begin{itemize}

\item \textit{Intra-organisational Distribution}. Distributed process in the same organisation face basic challenges in mobility and integrating the resource constrained participants. Although the model compiled code execution approaches \cite{Glombitza2011,Caracas2011,Casati2012,Tranquillini2012,Tranquillini2015} are promising and efficient, they lack standard approaches to convert the designed process models to machine-readable, executable program. Further, deploying the process model (either the raw model or the transformed version) to the edge nodes also involves challenges in performance, especially in a large-scale IoT environment where the events can occur quite often and the frequency of changing process is high.

\smallskip
\item \textit{Inter-organisational Cooperated Devices}. Collaborative IoT devices bring many new possibilities in IoT. Organisations can share their deployed IoT devices to one another in order to reduce the deployment cost. However, it will raise new challenges in privacy, trust, quality of service, service level agreement and negotiation. Currently, existing frameworks have not fully addressed these issues. Additionally, the inter-organisational BP model in IoT requires further study on the adaptive extension of the Public-to-Private workflow abstraction \cite{Aalst2001} in which the organisations need to design the common agreed high-level BP model standards.
\end{itemize}
\vspace*{-15pt}

\subsection{Challenges in Run and Adjust}
\subsubsection{Run-time Monitoring and Event Streaming}
As mentioned in the previous section, runtime monitoring in BPMS4IoT is a crucial task because it involves many factors that  relate to mobile computing and wireless sensor network domain. Generally, existing BPMS4IoT frameworks have not broadly studied about this topic. Although the time-out solution proposed by \cite{Dar2015} is capable of identifying the runtime failure of the edge network device, considering if the failed node is the cluster head of the edge network and there is no alternative node that can replace the cluster head, or if the failure is the current Internet provider of the edge network, the monitoring procedure of the edge network will fail entirely. Researchers in BPMS4IoT need to further investigate this issue.

Existing event streaming schemes \cite{Jung2012,Appel2014,Tranquillini2015} were designed for specific scenarios. There is a lack of generic BP model for defining the streaming type activities for both intra-organisational, inter-organisational and edge networks. In order to develop the generic streaming BP model, developers need to understand how the front-end devices integrate with the backend cloud services in the reliable channel and protocols. Further, they need to consider the performance and cost efficiency.

\subsubsection{Energy Efficiency} 
The large scale connected things provide various possibilities and also raise numerous challenges. One research interest in both academia and industry is in energy conservation. In general, IoT environment involves a large number of devices deployed in high density. These devices include battery-powered and AC-powered devices.

Besides the industrial standards such as CoAP and MQTT, which can reduce the energy consumption from data transmission, the application layer can also apply strategies such as utilising cloudlet-based architecture \cite{Gao2012} or utilising collaborative inter-organisational data brokering scheme \cite{Chang2015icws}. In the past, many works have discussed such collaborative network  environment \cite{Feamster2007,Sofia2008,Middleton2011,Frangoudis2011,Garikipati2013,Cao2015}, and the related commercial services already exist for many years (e.g. \url{http://www.fon.com}).

As the recent perspective in applying {\it software defined networking} (SDN) to the sensing cloud \cite{Distefano2015}, the inter-organisational collaboration is possible to be realised in higher level controlled by a software system instead of depending on the infrastructure hardware compliant. Hence, it leads to a research direction in developing the new BP model design that composes SDN, WSN and edge computing.

\subsubsection{Context-awareness}
There exists a number of literature surveys in context-aware workflow management \cite{Ardissono2007,Smanchat2008,Tang2008}, which can serve as the starting point for addressing context-awareness in BPMS4IoT. However, they are focusing on the classic systems in which no mobile IoT devices or the common wireless IoT devices involved in the workflow processes.

Commonly, a system can utilise the rule-based schemes mentioned previously in Section~\ref{sec:context-awareness} or utilise the machine-learning schemes based on Bayesian network, Markov chain to achieve context-awareness. In fact, recent research trends in  context-awareness have started applying process mining techniques \cite{Jaroucheh2011,Pileggi2015,Pileggi2015a}. Since process mining is a popular technique in BPM \cite{Dumas2013}, those process mining-based context recognition schemes can be quite promising in supporting context-aware BPMS4IoT.

\subsubsection{Scalability}
Besides the two topics mentioned previously in Section~\ref{sec:scalability} (i.e. the growing volume of stream data and the growing number of BP participative devices), scalability in BPMS4IoT also involves the growing number of location-based real-time applications, which is not a widely studied topic in BPMS, mainly because it is emerging when the recent information systems are highly relying on the distant data centres for all the processes. The location-based real-time applications such as urban AAL \cite{Chang2015scc} requires low latency response. However, the growing number of mobile users in the urban area increase the network traffic, in which the transmission speed of BP tasks (i.e. between the end-user application and the distant server) is hard to achieve the users' satisfaction. In order to resolve the issue, researchers in this domain can consider utilising a hybrid infrastructure that composes the distant mobile cloud-based BPMS4IoT with mobile edge cloud and mobile ad hoc cloud. Such a design further involves the BP model design and the challenge in runtime monitoring and on-demand reconfiguration.

\subsubsection{Intelligence}
Intelligence in BPMS4IoT involves challenges in efficient collecting and preprocessing data. These topics involve wise decision-making strategies supported either by the tools such as Complex Event Processing tool---open-source Esper engine used in edUFlow \cite{Jung2012} and as a part of the integration framework---LinkSmart used in EBBITS \cite{Furdik2013}, Tibco (\url{http://www.tibco.com}), Drools (\url{http://www.drools.org}) or other machine learning, data mining, artificial intelligence techniques. Since Cabanillas et al. \cite{Cabanillas2013} have discussed a list of major challenges in the event stream monitoring of BPMS for logistics that involves both cloud-side ERP system and the front-end vehicle-based mobile nodes, their work can serve as a guidance for the research in this direction.

Another major challenge of intelligence is the continuous optimisation in BPMS4IoT. In this topic, Process Mining and Process Discovery techniques show the promising solutions. For instance, Gerke et al. \cite{Gerke2009} have proposed a process mining algorithm to improve RFID/EPCGlobal-based logistic system. Zhang et al. \cite{Zhang2011} and Carolis et al. \cite{Carolis2015} have proposed the process mining schemes to improve AAL applications. All these schemes are directly related to BPMS4IoT. However, they were designed for specific application domains. Developers in different domains (e.g. mobile IoT, which involves mobility and unreliable connection issues) need further research to develop the most feasible solution for their systems.

\subsubsection{Big Data Service}
%
%
The deployment of BPMS4IoT will forward the enterprise systems to the Big Data era. In general, Big Data in IoT represents a vast volume of data generated from the IoT networks and organisations can make use of them, which is not available before. Ideally, organisations will gain benefit from the Big Data to improve and enhance their business process more efficiently and more intelligently \cite{Chandler2015}. In order to realise the vision, BPMS4IoT needs to address the challenge of {\it varying forms of data}. The data of IoT comes in various formats from different co-existing objects in IoT networks. The information system can utilise machine-learning mechanism to identify the correlation between the data from different objects and generate the meaningful information. However, processing the various formats of data may not be a swift task. For example, in order to identify a suspicious activity in an outdoor environment, the system may integrate the video data and temperature data from different sensors and then either utilise an external third party cloud service for the analysis processes or invoke the external database service to retrieve the related data and analyse them in the intra-organisational information system. The challenge is if such a need is on-demand, how does the system generate the result in time because it involves the data transmission time in different networks and the large volume of data processing.

Research in addressing Big Data of BPMS4IoT can further refer the study in \cite{Munoz-Gama2014,Aalst2015}, which provide a guidance in identifying the requirement of this domain. The works have introduced a new concept---{\it Internet of Events (IoE)}, which represents a large volume of event stream data coming from IoT system, and what are the promising process mining techniques can overcome the issues in IoE.

\section{Conclusion and Future Directions}

The Workflow Management Coalition introduced the notion---{\it BPM Everywhere} \cite{Fischer2015} to represent the future IoT in which almost any part of the IoT system will utilise BPM. In such an IoT system, the deployment of BPMS is not only in the intra-organisational information systems but it also composes the inter-organisational BP activities. Specifically, the entities involved in the system include both back-end cloud services and the front-end edge network established by clusters of interconnected IoT devices and also the Fog service nodes. In order to realise such a vision, BPMS4IoT will face many specific challenges in each of their life cycle phase, which were not fully addressed in the past BPMS. 

\begin{figure*}[h]
  \centering
    \includegraphics[width=0.65\textwidth]{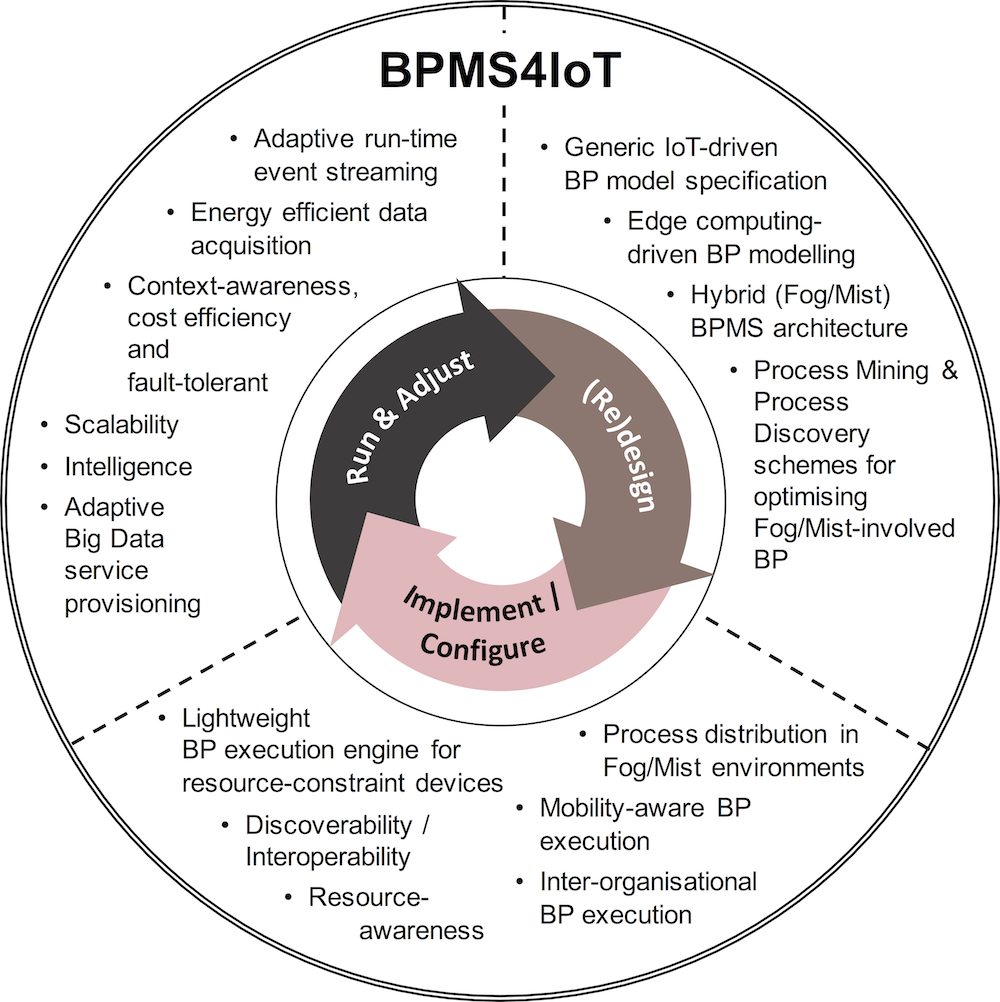}
    \caption{Research roadmap in BPMS for IoT.}
    \label{fig_roadmap}
\end{figure*}

This paper has discussed the state-of-the-art and challenges involved in each lifecycle phase of BPMS. The study has shown that existing frameworks have not yet addressed many research challenges involved in BPMS4IoT. Further, most of the challenges highly relate to the research in MCC domain, which raises the opportunity to MCC discipline.~In summary, Figure~\ref{fig_roadmap} illustrates a research roadmap as a future research direction in Mobile Cloud-based BPMS4IoT.

\bibliographystyle{elsarticle-num}
\bibliography{Ref.bib}

\end{document}